\documentclass[12pt,preprint]{aastex}
\usepackage{emulateapj5}

\usepackage{epsfig}
\usepackage{graphicx}
\def\gta{\mathrel{\hbox{\rlap{\hbox{\lower4pt\hbox{$\sim$}}}\hbox{$>$}}}}

\shorttitle{{\it Spitzer} Virgo ICL}
\shortauthors{Krick et al.}

\begin{document}
\newcommand\msun{\hbox{M$_{\odot}$}}
\newcommand\lsun{\hbox{L$_{\odot}$}}
\newcommand\magarc{mag arcsec$^{-2}$}
\newcommand\h{$h_{70}^{-1}$}

\bibliographystyle{apj}
\title{\bf Spitzer IRAC Low Surface Brightness Observations of the Virgo Cluster}

\author{J.E.~Krick\altaffilmark{1}, C.~Bridge\altaffilmark{2},
  V.~Desai\altaffilmark{1}, J.C..~Mihos\altaffilmark{3},
  E.~Murphy\altaffilmark{1}, C.~Rudick\altaffilmark{4}, J.~Surace\altaffilmark{1}, and J.~Neill\altaffilmark{2}}

\altaffiltext{1}{Spitzer Science Center, MS 220--6,
California Institute of Technology, Jet Propulsion Laboratory,
Pasadena, CA 91125, USA}
\altaffiltext {2}{Division of Physics, Math, \& Astronomy, California Institute of Technology, Pasadena, CA 91125}
\altaffiltext {3}{Department of Astronomy, Case Western Reserve University, 10900 Euclid Ave, Cleveland, OH, 44106}
\altaffiltext {4}{ETH Zurich, Institute of Astronomy, CH8093 Zurich, Switzerland}
\email{jkrick@caltech.edu}

\begin{abstract}

  We present 3.6 and 4.5 \micron~{\it Spitzer} IRAC imaging over 0.77
  square degrees at the Virgo cluster core for the purpose of
  understanding the formation mechanisms of the low surface brightness
  intracluster light features.  Instrumental and astrophysical
  backgrounds that are hundreds of times higher than the signal were
  carefully characterized and removed.  We examine both intracluster
  light plumes as well as the outer halo of the giant elliptical M87.
  For two intracluster light plumes, we use optical colors to
  constrain their ages to be greater than 3 \& 5 Gyr, respectively.
  Upper limits on the IRAC fluxes constrain the upper limits to the
  masses, and optical detections constrain the lower limits to the
  masses.  In this first measurement of mass of intracluster light
  plumes we find masses in the range of $5.5\times 10^8 - 4.5\times
  10^9$ and $2.1\times 10^8 - 1.5\times 10^9$ \msun~for the two plumes
  for which we have coverage.  Given their expected short lifetimes,
  and a constant production rate for these types of streams,
  integrated over Virgo's lifetime, they can account for the total ICL
  content of the cluster implying that we do not need to invoke ICL
  formation mechanisms other than gravitational mechanisms leading to
  bright plumes.  We also examined the outer halo of the giant
  elliptical M87.  The color profile from the inner to outer halo of
  M87 (160 Kpc) is consistent with either a flat or optically blue
  gradient, where a blue gradient could be due to younger or lower
  metallicity stars at larger radii.  The similarity of the age
  predicted by both the infrared and optical colors ($>$ few Gyr)
  indicates that the optical measurements are not strongly affected by
  dust extinction.

\end{abstract}


\section{Introduction} 
\label{intro}
The basic picture of hierarchical structure formation is that large
galaxies assemble from smaller pieces.  A galaxy's environment during
assembly is expected to play a significant role in its evolution.
Indeed, galaxies in the dense cores of clusters differ systematically
from those in the coeval field, in terms of their morphologies,
stellar populations, and gas content
\citep[e.g.][]{dressler1997,butcher1978,butcher1984,desai2007,krick2009}.

Many different physical processes have been implicated in driving the
evolution of cluster galaxies, including both gravitational and gas
dynamical effects. Gravitational effects include slow galaxy-galaxy
interactions and mergers \citep{mihos2004}, harassment
\citep{moore1996,moore1998}, the effects of the global tidal field
\citep[e.g.][]{Byrd1990,henriksen1996} and the effects of cluster
sub-cluster merging \citep{bekki1999,stevens1999}.  Gas dynamical
effects include ram pressure and turbulent viscous stripping of the
cold ISM \citep{gunn1972,abadi1999,schulz2001,vollmer2001,murphy2009} and
starvation \citep{Larson1980}. 

Within a cluster, gravitational effects can lead to the removal of
stars from their parent galaxies into the intracluster medium (ICM).
Gas dynamical effects can additionally strip gas and dust from
galaxies.  This stripped material may subsequently become sites of
star formation within the ICM. The properties of these stars, as
traced by the intracluster light (ICL), can constrain the types and
frequency of the physical processes at work in clusters
\citep{krick2006,krick2007,gonzalez2005,feldmeier2004,zibetti2005}
by serving as a fossil record of the interaction history of
the cluster.

Large stellar streams from galaxy interactions are a gravitational
effect we know occurs in clusters.  Massive streams will be visible
for of order $\sim 1-2$ times their dynamical time before dispersing into the
diffuse ICL population \citep{rudick2009}.  If the streams can account for all of the
diffuse ICL built up over time, then there is no need to invoke
other mechanisms such as gas dynamical effects including ram pressure
stripping or starvation in generating large amounts of ICL.

Virgo is one of the best places to study these evolutionary processes
in action.  First, Virgo is the only cluster that is close enough
\citep[16.6~Mpc;][]{shapley2001} to easily achieve many resolution
elements per square kiloparsec.  Second, many of its galaxies have
clearly been modified by their environment
\citep{2011arXiv1101.4066A}.  Third, the cluster exhibits substantial
substructure, with infalling groups containing tidally interacting
galaxies and mergers, as seen for the M87 \& M86 regions
\citep[e.g.][]{schindler1999,kenney2008}.  Finally, Virgo is
well-studied at a range of wavelengths (there are at least 677
literature references in NED for the Virgo cluster).

As a central dominant (cD) galaxy of the Virgo cluster, M87 is a
prototypical giant elliptical. Its optical surface brightness profile
is poorly fit by a single deV profile; instead, better fits are
achieved through high-n Sersic profiles \citep{kormendy2009, janowiecki2010} or a combination of single deV profile with an
additional outer component (Liu et al 2005). Either profile shape
implies a significant amount of luminosity at large radius, possibly
attributable to an ICL component.  M87 is known to have a flat to
slightly blue color gradient from the inside to outer regions of the
galaxy \citep{liu2005, zeilinger1993}. A flat color profile implies a
constant age as a function of radius.  A color gradient, even in the
near infrared, would imply changing age as a function of radius.  Flat
age profiles cannot rule out monolithic collapse scenarios for the
formation of this giant elliptical .

At present, the only cluster-wide information obtained for the distribution of ICL
in Virgo comes from the optical ($V \& B$-band) observations of
\citet{mihos2005} and \citet{rudick2010}.  These data reveal an
intricate complex of intracluster stellar features, including long ($>
100$kpc) streamers, arcs, tails, bridges, and diffuse blobs
unassociated with galaxies.  These data are clear evidence for
interactions between the infalling populations of smaller galaxies
with the cluster cD (M87) and the infalling M86 sub-cluster structure.

In this paper we present a map of the central 0.77 square degrees in Virgo using
IRAC at 3.6 and 4.5 \micron.  These wavelengths are interesting
because they 1) are sensitive estimators of stellar mass,
including that contributed by old stars; 2) will provide a color
measurement of the diffuse features in this cluster; and 3) are
largely unaffected by the presence of dust.

This paper is structured in the following manner. In
\S\ref{observations} and \S\ref{data} we discuss the data including
specific reduction techniques for low surface brightness (LSB)
measurements with {\it Spitzer}.  Detection limits are described in
\S\ref{noise}.  \S\ref{results} discusses the
results for the intracluster plumes and halo of M87.  The paper is
summarized and conclusions are drawn in \S\ref{conclusion}.
Throughout this paper we use $H_0=71$km/s/Mpc and $\Omega_M$ = 0.27.
With this cosmology, we use the \citet{shapley2001} luminosity
distance of 16.6 Mpc corresponding to a redshift of 0.0039.


\section{Spitzer Warm IRAC Observations }
\label{observations}

\begin{figure}
\epsscale{0.6}
\plotone{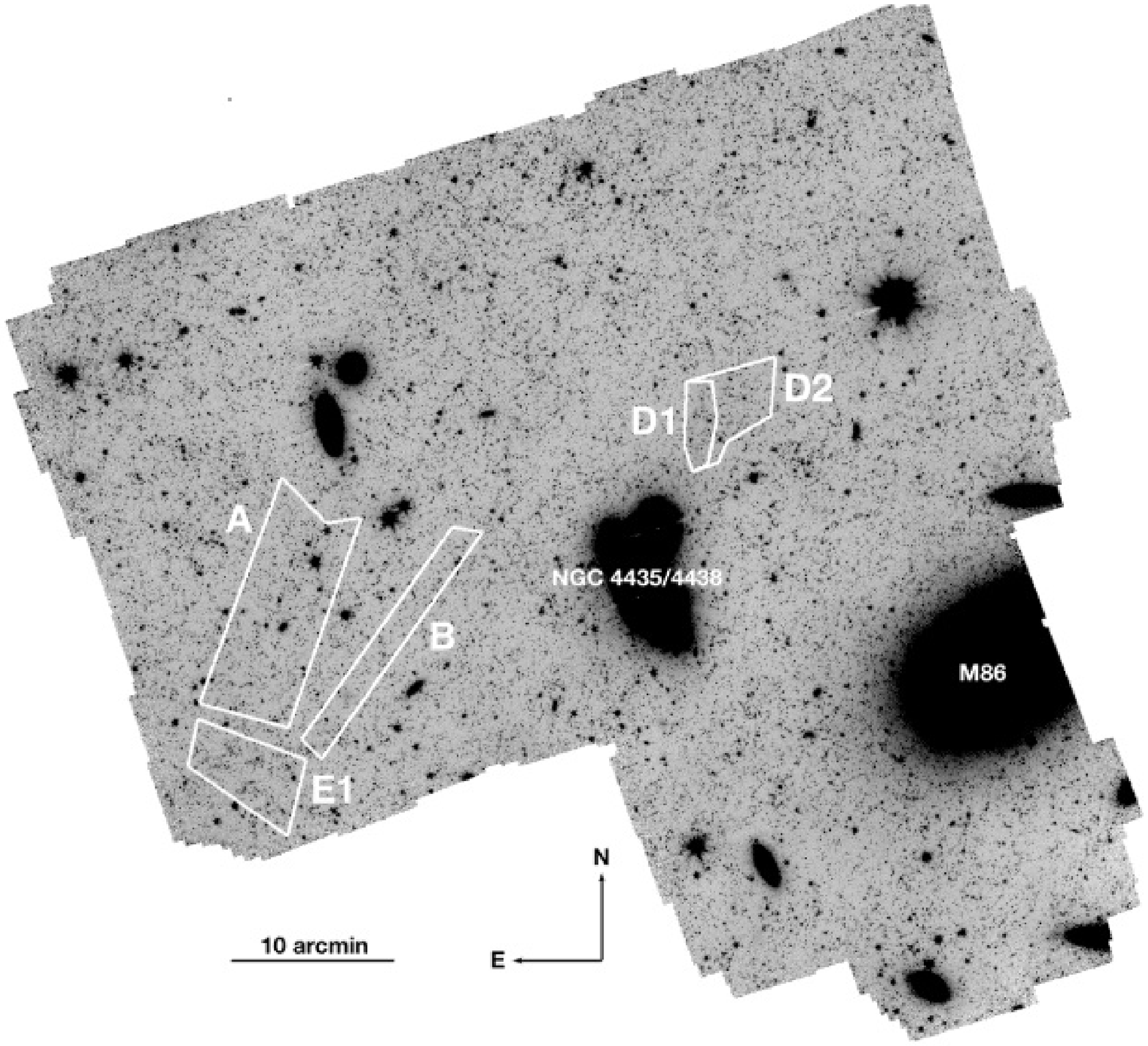}
\epsscale{0.5}
\plotone{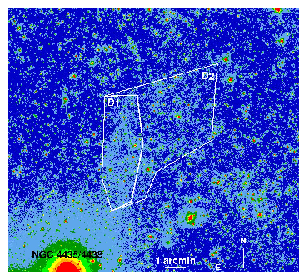}
\caption[ch1]{Top: The full 3.6 \micron~ mosaic, including
  intracluster and M87 halo regions as used in the
  text. Bottom: A zoom in on plume D where the low surface brightness
  signal is visible in the D1 region.}
\label{fig:ch1}
\epsscale{1}
\end{figure}

We make a continuous map of the central 0.77 square degrees of the
Virgo cluster between M87 and M86 where many optical intracluster
light features have been identified \citep[][see section
\S\ref{plumes} and Figure \ref{fig:ch1}]{mihos2005,rudick2009}.  We
designed our observing program to reach the lowest possible surface
brightness in the allotted time (100 hours; PID 60173).  Our expected
depth, based on the {\it Spitzer} sensitivity estimator, is 0.0002
Mjy/sr at 3.6 \micron.  Discrepancies from this will be discussed below.  An
additional design constraint was that the map include an ``off''
cluster region for measuring the non-cluster background level since we
expect the diffuse intracluster light to be prevalent.

Warm IRAC \citep{fazio2004} simultaneously observes with two detectors
at 3.6 \& 4.5~\micron~ (channel 1 \& 2 respectively). Each detector has
a $5\farcm2 \times 5\farcm2$ field of view with 1\farcs2
pixels. Images are acquired in two adjacent fields, separated by
6\farcm5.  Our entire region is imaged in five layers of observations,
each 10 min per pixel or 50 min per pixel total over all layers.
Inside of each layer we dither inside of a mapping pattern to cover
the entire area while facilitating the removal of cosmic rays.  We use
a 100~sec frame time which gives a coverage of at least 30 overlapping
frames per pixel.  3.6 \& 4.5 \micron~ data were taken simultaneously
and so have similar coverage, except for one array width on the north
and south side of the final mosaic.  Data were taken during two
consecutive, two week long, IRAC campaigns PC15 \& PC16 during
February, 2010. Due to scheduling constraints, the AORs are spread out
in time between these two campaigns, i.e. they are not taken
continuously in time.


\section{Data Reduction}
\label{data}

Data reduction to reach our intended low surface brightness detection
limits is a challenge because it is a factor of hundreds below the
background level in the images and very little work has been done to
characterize the behavior of the IRAC instrument at these levels.  The
following sections discuss our data processing steps with an eye
towards the errors in each step.  We start with a standard data
reduction for both channels.  We use the {\it Spitzer} Science Center
(SSC) pipeline which generates corrected basic calibrated data files
(CBCD; pipeline version S18.18) for our data reduction.  These data
have already been dark and flat-field corrected with standard
calibration darks and flat-fields.  A flux conversion has already been
applied based on A \& K type standard stars placing the data in units
of MJy/sr.

\subsection{Dark \& Bias Correction}

Understanding the dark and bias contribution is important for LSB
observations as it is a significant fraction of the total counts in
each pixel.  Calibration darks relevant to this dataset are taken as a
set of 28 dithered 100s exposures within seven days of our
observations.  Standard darks can't be taken with IRAC because the
shutter is not ever used.  The darks are taken by looking at a region of
very low background near the north ecliptic pole.  Those images are
then median combined to remove sources which leaves an image of both
the bias and dark current.  Upon subtraction from the science data,
both of these effects are removed. Because the dark field does not
have a completely zero celestial background, this dark removal
technique will also remove some real celestial background.  We account
for this in our program design by intentionally observing an ``off''
background location in the cluster which we will define to be of zero
background, and tie the rest of the image to that (see
\S\ref{mosaic}).

\subsection{Flat Field Correction}
The accuracy of any LSB measurement is limited by fluctuations on the
background level. A major contributor to those fluctuations is the
large-scale flat-fielding accuracy.  The IRAC flat-field is derived
from dithered observations,accumulated over 110 hours, of the zodiacal
background as the best approximation to a source which uniformly and
completely fills the field of view.  The pixel-wise accuracy
of the flat-fields is very high, 0.2\% \& 0.1\% at 3.6 \& 4.5 \micron~
respectively. There is no variation with time in the flat field to the
levels quoted.



\subsection{Illumination Correction}
\label{illuminate}
There remains an illumination pattern in the science data in part due
to latent images.  Bright, saturated sources on prior images can fill
traps in the arrays which decay away at a rate that is longer than the
time between exposures thus leaving apparent flux at the location of
that bright source in a previous image.  Latent images can also result
from the telescope slewing across a bright star which will leave a
stripe of populated traps across the image.  We removed these latents
from the data by making a tailored illumination correction for each
set of observations.  In each individual CBCD, we mask all the
detected sources based on a SExtractor \citep{bertin1996} segmentation
map.  SExtractor is run with a 1.5 sigma detection and analysis
threshold and a minimum of 5 pixels for object detection.  Then, the
masked images are median combined inside of each observing set to
create the illumination correction.  Median combining these masked
images does an excellent job of creating an illumination correction
because there are of order 100 dithered images per channel.  Because
we only want to remove the pattern in this illumination image, and not
the background level, we first subtract the mean level from the
illumination image, then we subtract the illumination pattern from the
science data.


\subsection{Zodiacal Light Removal}

The dominant, non-instrumental background source in space-based
infrared observations is the zodiacal light.  This is light scattered
from interplanetary dust and depends on both time of year and
direction of observation.  The zodiacal light is assumed to be uniform
on the size scale of the IRAC field of view.  A single value of the
zodiacal light per image is removed from the images using a model
based on COBE DIRBE data \citep{kelsall1998} as listed in the headers.
This model is expected to be accurate to 2\% or better.  Values for
the zodiacal light are in the range of 0.104 to 0.114 MJy/sr at 3.6
\micron~ and 0.238 to 0.323 MJy/sr at 4.5 \micron.  Zodiacal light
values are listed in the headers in units of MJy/sr.  However the
conversion from data units to MJy/sr provided by the {\it Spitzer}
Science Center is based on point sources.  Because the zodiacal light
is an extended source, this point source derived conversion is not
appropriate and we apply a multiplicative, extended source aperture
correction to the zodiacal light value (0.91 and 0.94 at 3.6 \& 4.5
\micron~respectively) before subtraction.



\subsection{First Frame Effect}

\begin{figure}
\begin{center}
\includegraphics[angle=90, scale=0.5]{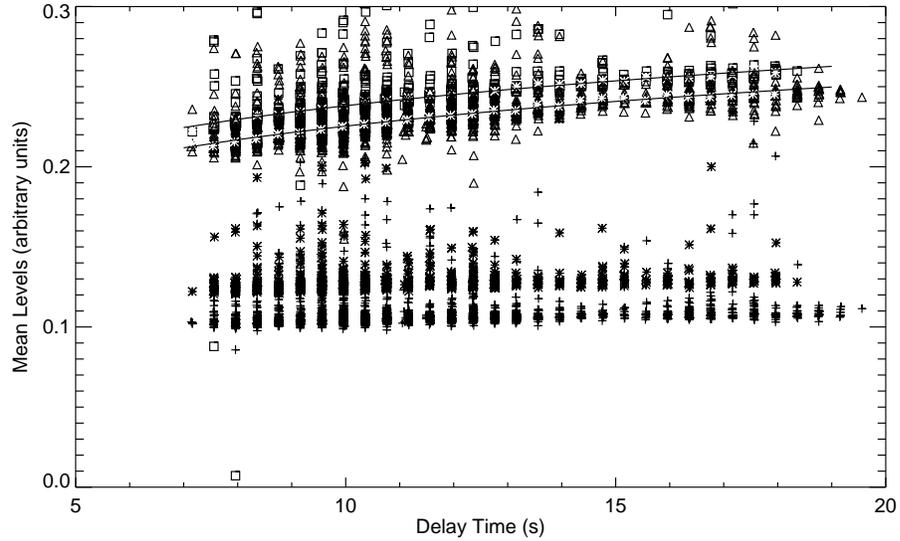}
\end{center}
\caption[check_ffcorr]{First Frame Correction: All $\sim 3000$ images
  at both 3.6 (squares+triangles) and 4.5 \micron~ (asterisks +
  pluses) are shown in this plot of delay time vs. mean background
  level.  Mean levels have been arbitrarily scaled to show both
  channels on the same plot.  The data from the two campaigns have
  different mean background levels as shown here with the different
  symbols within each channel.  For the 3.6 \micron~ data the mean value in each
  bin of delay time is shown with the white asterisks which is then
  fitted with a logarithmic function shown by the solid lines.  }
\label{fig:check_ffcorr}
\epsscale{1}
\end{figure}

The IRAC arrays are known to have a relation between the bias level
and the delay time between frames. This is referred to in the
documentation as the ``first frame effect'', even though it effects
{\it all} frames.  Here, delay time refers to the time between one
exposure and the next, and not the exposure time itself.  There are
different delay times between our images which all have the same
exposure time because of the dithering and tiling pattern we have
chosen.  Figure \ref{fig:check_ffcorr} shows this effect for our
dataset.  Specifically, 3.6 \micron~data shows a logarithmic trend between
bias level and delay time.  The mean level shown is measured from
masked individual CBCD frames that are column pulldown corrected.
Masks are derived from SExtractor segmentation images.  We fit the
first frame effect separately for PC15 and PC16 with a logarithmic
function and apply this as a correction to the data.

There is no measurable first frame effect in our 4.5 \micron~ data (see
Figure \ref{fig:check_ffcorr}).  This is consistent with the known
behavior of the arrays.

\subsection{Clock Time Effect}

\begin{figure}
\epsscale{0.4}
\plotone{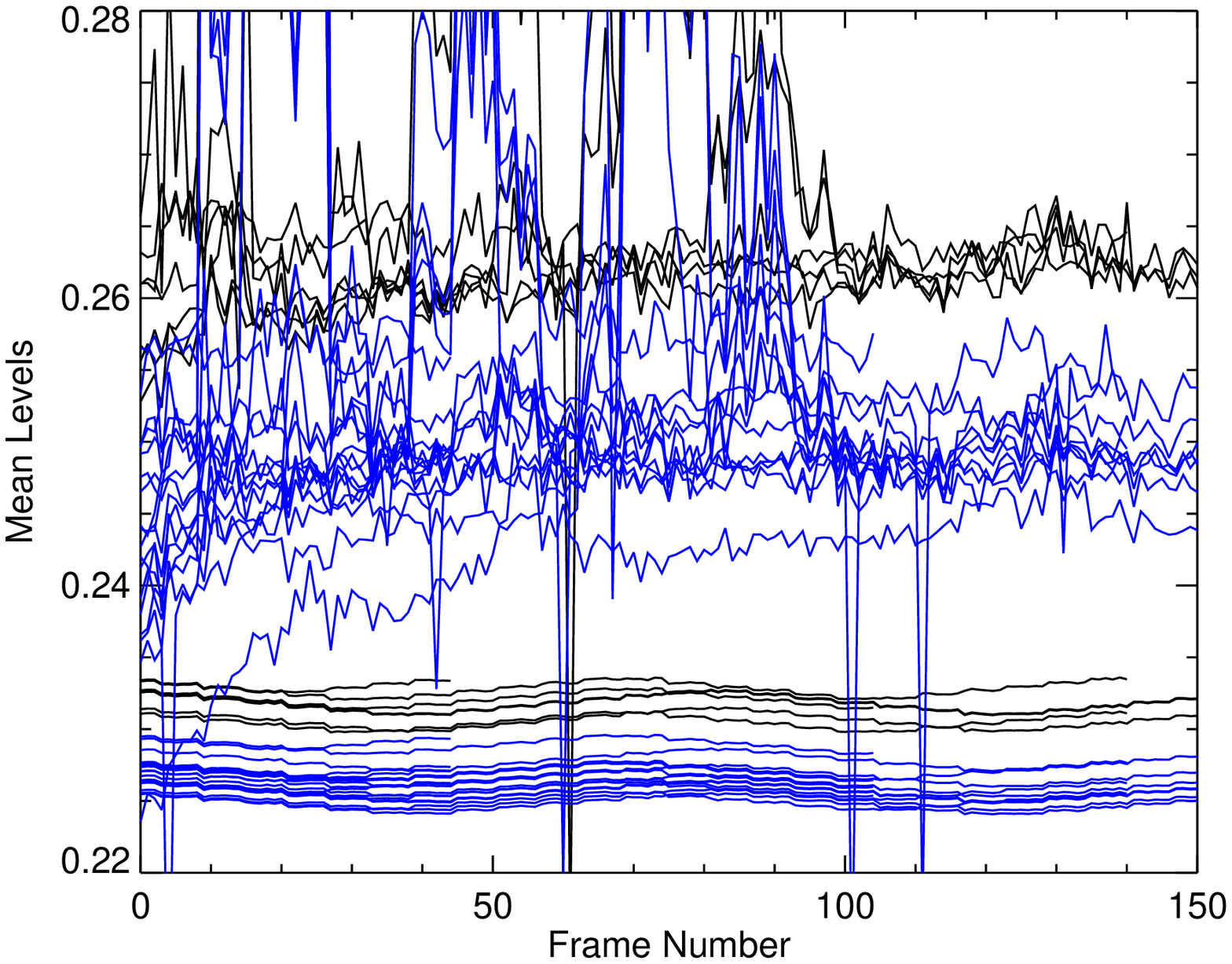}
\plotone{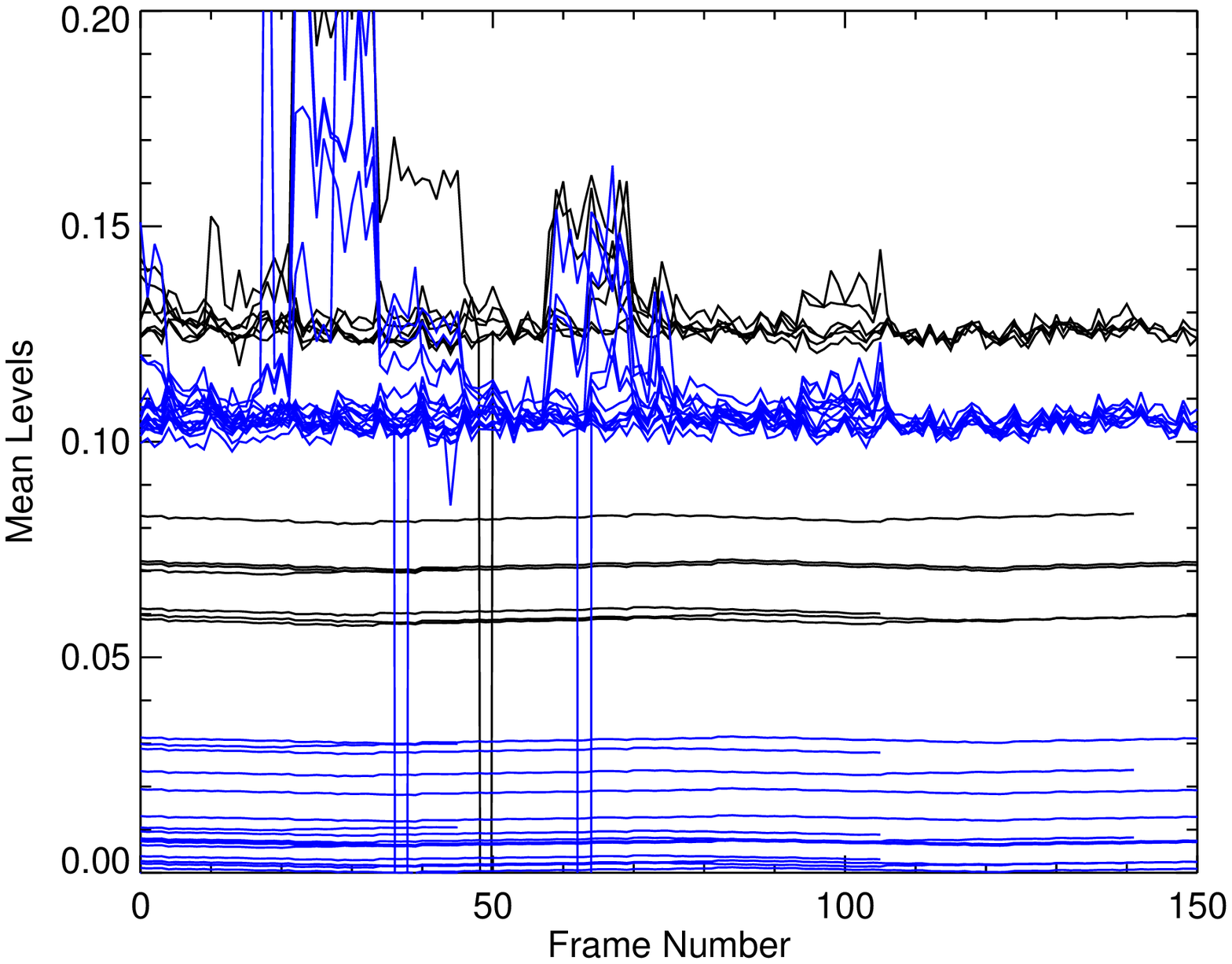}
\caption[meanvstime_ch1]{Relaxation effect for 3.6 \micron~(left plot)
  and 4.5 \micron~(right plot): Each of the 25 AOR's is shown in this
  plot of frame number within the AOR vs. mean background level
  (arbitrary units) as solidly squiggly lines.  AORs within campaign
  pc15 are shown in black, and those in campaign pc16 are shown in
  blue.  Zodiacal light model levels are shown as smooth lines in the
  bottom third of the plot, arbitrarily scaled to fit in the frame.
  Spikes in mean level occur when a bright star or galaxy is on the
  frame.  Our cycling dither pattern exposes some objects on multiple
  consecutive frames leading to spikes wider than one frame number.}
\label{fig:meanvstime_ch1}
\epsscale{1}
\end{figure}

After removal of the first frame effect, we notice a secondary,
undocumented effect between mean levels and clock time at 3.6 \micron.
Figure \ref{fig:meanvstime_ch1} shows mean levels versus frame number
for each of the 25 AORs as well as the zodiacal light model pattern
for the same frames.  Spikes in this plot occur when a bright star or
bright galaxy are in the frame.  There are two correlations: (1) a
ramping in mean level from the beginning of the AOR to the end, and
(2) mean level variation from AOR to AOR.  We see a large jump in
background levels between campaign PC15 data and campaign PC16 data,
but also small level changes within the campaigns.

The zodiacal light pattern is not likely to be responsible for the
ramping effect.  This is demonstrated in the lower portions of 
Figure \ref{fig:meanvstime_ch1} where the zodiacal light contribution
is shown.  It has a significantly different shape than the ramping
effect.  Neither the jump or the ramp are correlated with position on
the sky or time since dark frames were taken.  This rules out
astrophysical background and build-up of flux on the chip as causes of
the effect.

It is likely that these effects are caused by the previous observation
history of the chip.  The ramping from the beginning to end of the AOR
is likely a relaxation in the chip. As these AORs were not all taken
sequentially in time, there is no way of knowing the precise
observation history of the chips, and therefore no obvious
correlations can be drawn with time since bright star observations or
high background observations.  The cause of the jump in background
levels is unknown although potentially caused by zodiacal light model
inaccuracies or previous observation history as with the ramping.

Because these effects can not be extragalactic in origin, we have
chosen to remove them from the data.  Each AOR trend is fit with
an exponential function which is then subtracted from the data.  We
choose an exponential function as the appropriate model for a
relaxation effect which fits the data well.

At 4.5 \micron~ there is the level change effect although not the ramping
seen at 3.6 \micron~ (see Figure \ref{fig:meanvstime_ch1}).  Again, we
remove the level change from the data to put all AOR's on the same
background level.



\subsection{The Mosaic \& Background Determination}
\label{mosaic}
After removal of the above discussed instrumental and zodiacal light
effects, we use MOPEX \citep{makovoz2005} for the final astrometry
alignment and mosaicing, see Figure \ref{fig:ch1}. M86 is marked for reference.
M87 is in the lower left corner, not imaged in our survey.


We determine a global background level from the northern edge of the
mosaic as an off-cluster region so that we can have an absolute
measure of ICL flux in cluster regions.  This assumes we have taken
care of all instrumental and local celestial effects, and the only
remaining source of flux in the background regions is cluster
light. This corner of both the optical and IRAC images appears free of
low surface brightness cluster flux.  Of course there still could be
some cluster flux at this location, so any measurement of the diffuse
cluster light from this data will be a lower limit.  This spot has no
saturated stars, is 30 arcminutes from M86, and 1.5 degrees from M87.
Before measuring the background, we use SExtractor to make a
segmentation image.  All sources are masked from the region.  We also
include only those pixels with a coverage of at least 25 100s images.
The background level determined in this corner is subtracted from the
entire image to reach our final background level.

\subsection{The IRAC PSF}
\label{psf}

\begin{figure}
\plotone{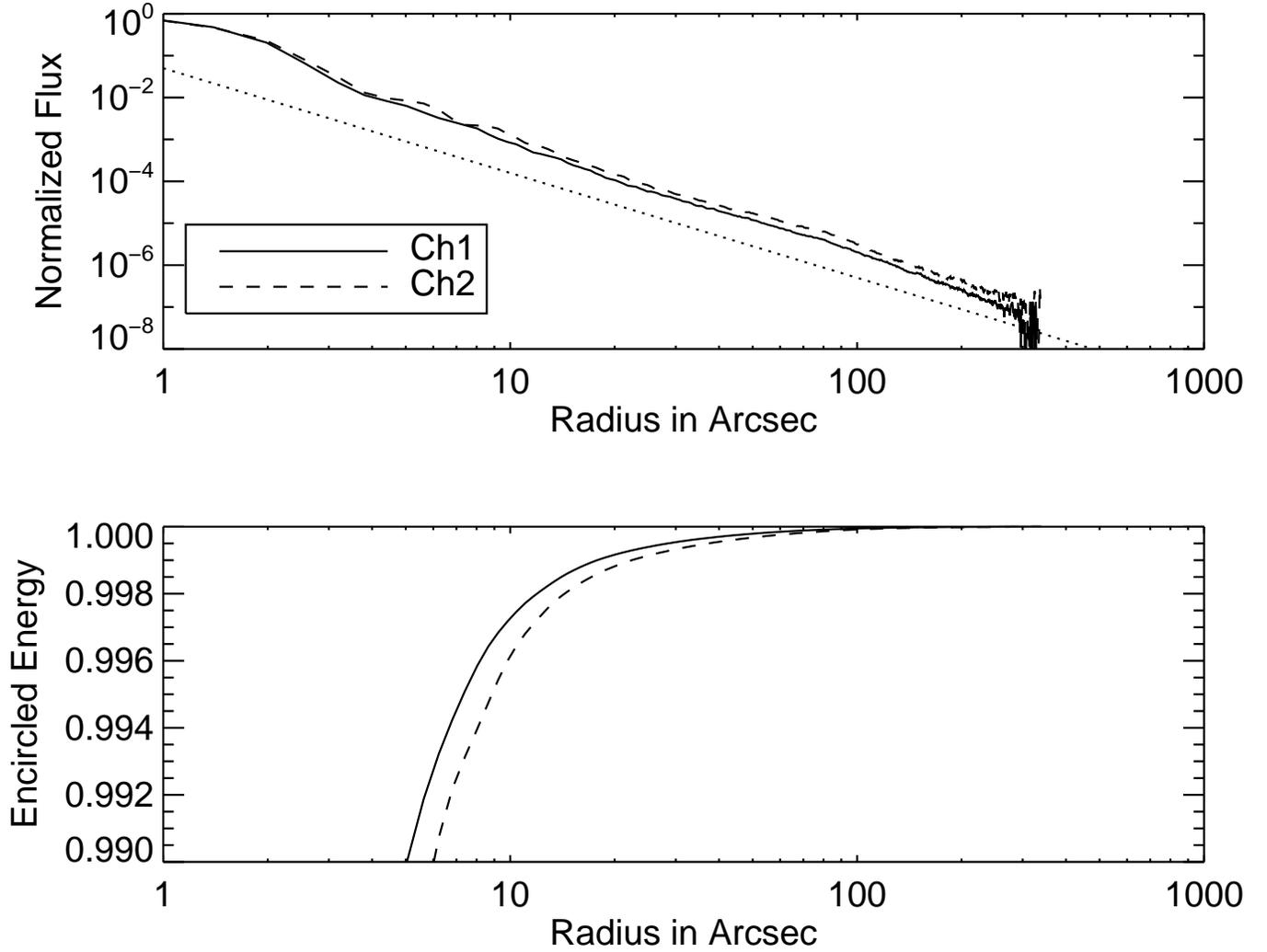}
\caption[psf]{Top: IRAC PSF to 300 arcseconds.  3.6 \& 4.5 \micron~ are shown in solid and dashed lines respectively.  The dotted line
  shows an $r^{-2.5}$ curve for reference.  Bottom: Encircled energy
  of the above IRAC PSF, here showing only the last 1\% of the stars
  flux.  The IRAC PSF is well behaved and does not push a lot of flux
  to large radii.  }
\label{fig:psf}
\epsscale{1}
\end{figure}

Because our images are plagued by source confusion, we study the shape
of the PSF to assure that there is not a lot of flux at large radii
from all the objects which could overlap to mimic a LSB signal.  We
use a very bright, saturated star (K$_{2mass}$ = 5.5 mag.) in our
mosaic to trace the outer wings of the PSF with high signal to noise,
see Figure \ref{fig:psf}.  This profile is made by first masking all
objects in the frame, {\it except} the bright star, to twice their
SExtractor measured isophotal radii.  We then measure the median flux
in radial bins out to $\sim 300$ arcseconds.  For reference the dotted
line shows an $r^{-2.5}$ profile offset from the data for ease of
viewing.  Because the core of this star has been corrected for
saturation, we can turn the radial profile into an encircled energy
profile by integrating under the data curve.  We find that 99.7\%
(99.6\%) of the star's flux is contained within 10 arcseconds of the
central pixel at 3.6 \micron~ (4.5 \micron).  There is not a large amount
of flux pushed to large radii by the PSF.






\section{Detection Limits}
\label{noise}
\begin{figure}
\epsscale{0.4}
\begin{center}
\includegraphics[angle=90, scale=0.5]{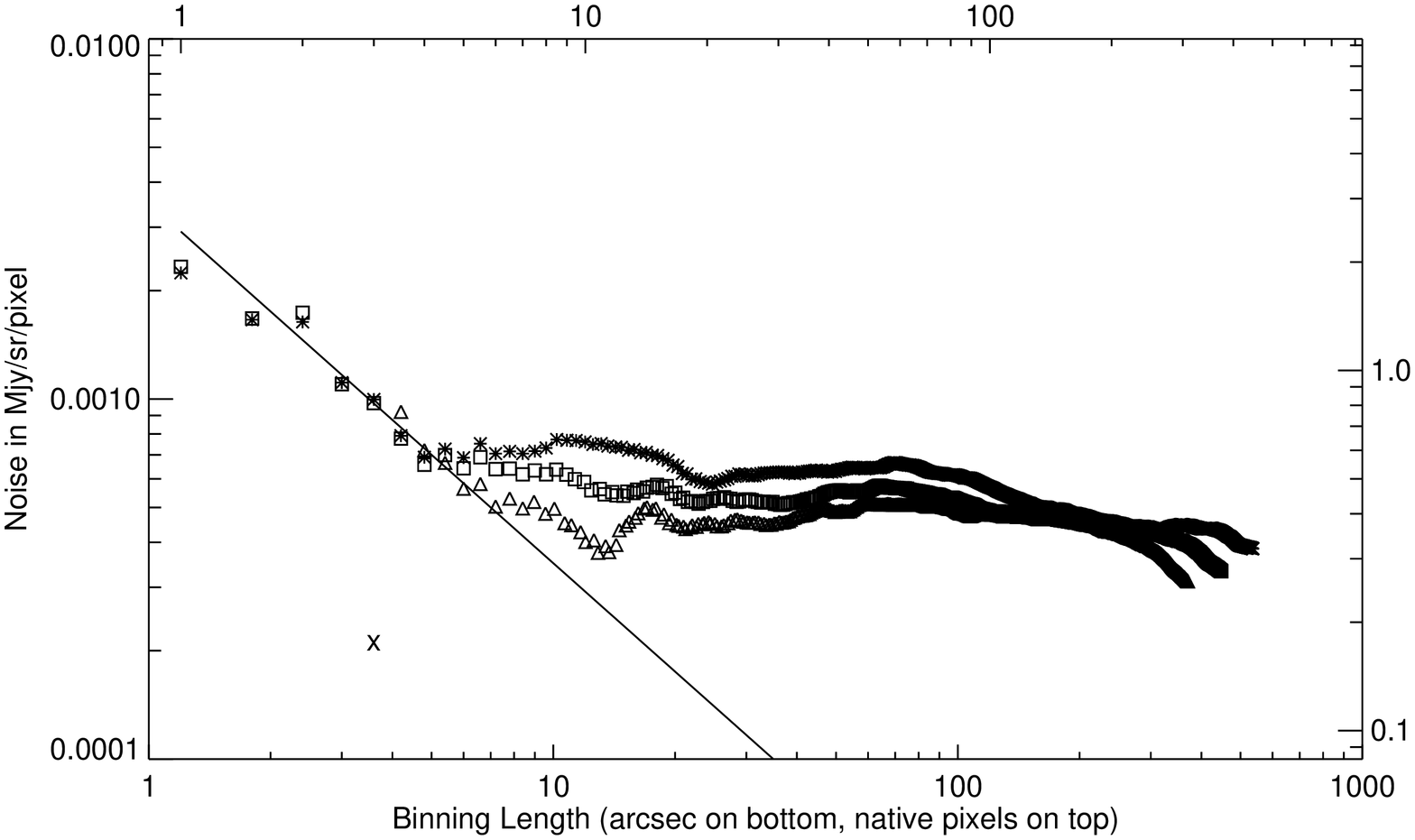}\\
\includegraphics[angle=90, scale=0.5]{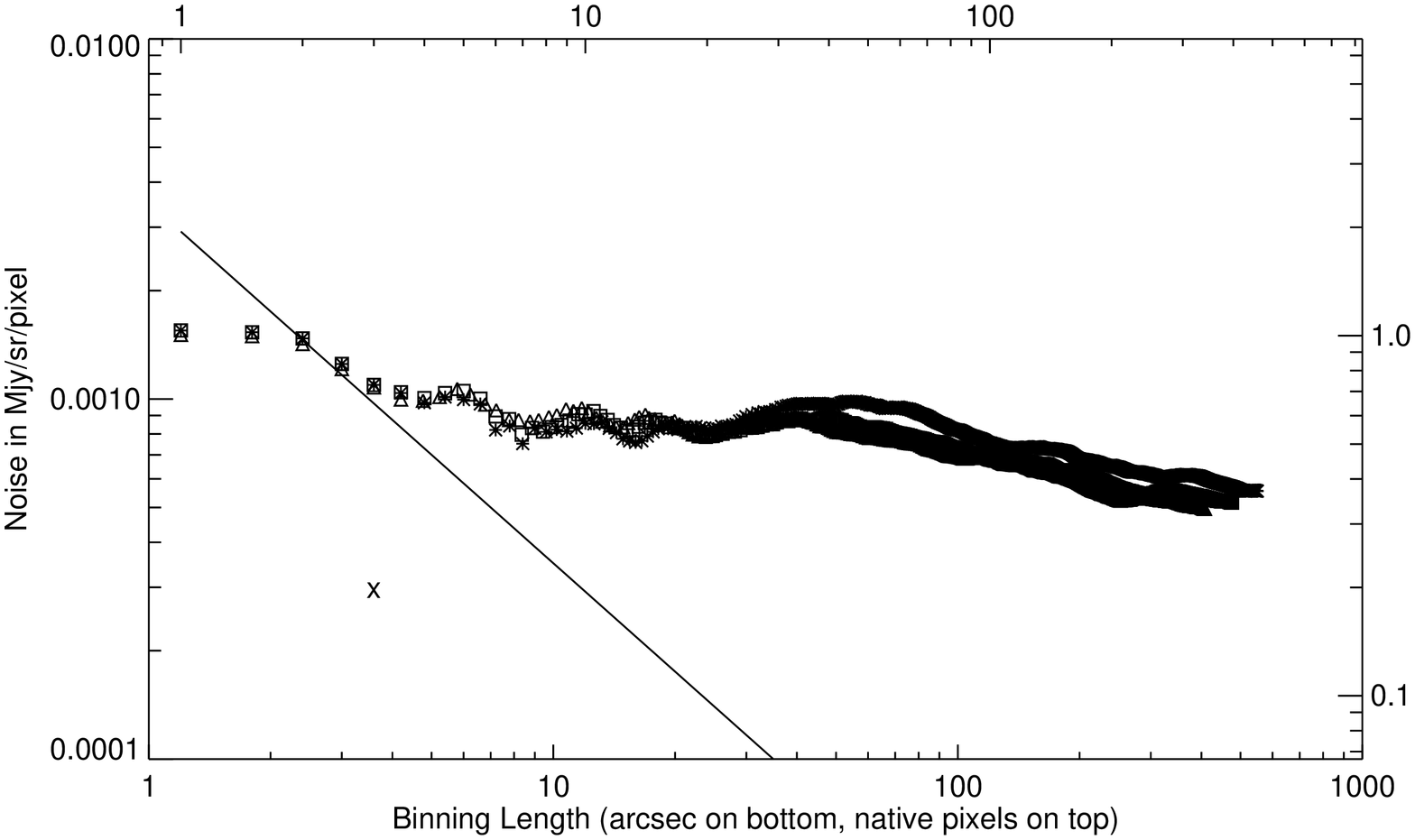}
\end{center}
\caption[binning]{Noise properties of the mosaic; 3.6 \micron~ on top,
  4.5 \micron~ on bottom.  For reference, the right hand y-axis is in units of
  nw/m$^2$/sr used in extragalactic background light studies.
  Asterisk symbols are for simple object masking, squares and
  triangles show masking to 1.5 and 2.0 times the object radii.  The
  'x' marks the sensitivity we expected to reach based on the {\it
    Spitzer} online performance estimation tool.  The solid line shows
  a linear relation for reference.  While object masking does bring
  the noise levels down, we still do not reach the levels expected
  from binning implying further noise sources creating a noise floor
  at $\sim 0.0005 \& 0.0008 $ MJy/sr/pixel for 3.6 \& 4.5 \micron~
  respectively.}
\label{fig:binning}
\epsscale{1}
\end{figure}

In order to observe intracluster light features we bin our mosaiced
data to achieve lower surface brightnesses.  Binning the data should
reduce the noise in the image linearly with binning size.  In Figure
\ref{fig:binning} we show a plot of noise versus binning size for the
data.  We measured the surface brightness in nine (six for 4.5
\micron) regions of one pixel by one pixel across the SExtractor
masked mosaic which are not near bright galaxies, stars, or optically
detected ICL features.  The noise is taken to be the standard
deviation of those nine (six) regions.  We then increase the size of
the box incrementally, measuring the standard deviation at each step,
until the box length is many hundreds of pixels.  We are unable to
measure this for a larger number of regions because of these large
final sizes and the available ``empty'' space in the image.
Specifically the 4.5 \& 3.6\micron~images have different final shapes
and so different numbers of blank regions.  The actual length of the
box is calculated from the square root of the number of unmasked
pixels.  For reference the solid line in Figure \ref{fig:binning}
shows a linear relation.  The 'x' marks the sensitivity we expected to
reach based on calculations with the {\it Spitzer} online estimation
tool.  The 3.6 \& 4.5 \micron~ mask shapes are calculated separately
from each image to insure that the masking is consistently done to the
same surface brightness threshold.  The measured noise does not
achieve the expected linear relation with binning length (see the
Appendix for further discussion of how to reach the lowest possible
surface brightness limits).

We believe that this technique for measuring the noise is
conservative.  Ideally we would like to measure the noise from many
regions in close proximity to the plume in question which do not have
any diffuse ICL in them.  This is simply not possible as the diffuse
intracluster light is ubiquitous and the area in the images limited.

\subsection{Small Scales}

There is a discrepancy at short binning length scales of just a few
pixels.  This discrepancy occurs because we are measuring noise on a
mosaiced image so on small pixel scales ($\sim$ few pixels), the noise
is correlated and will not bin down appropriately.  We have confirmed
that on a single CBCD image, the noise on small scales does behave
linearly with binning length down to a resolution element.

\subsection{Medium Scales}
\label{noise_medium}
On scales from five to 30 arcseconds much of the extra noise is due to
sources in the image.  This is demonstrated by changing the size of
the masks.  The first level of masking is just to use the SExtractor
segmentation map as as mask (see \S \ref{illuminate} for a description
of parameters used).  The resulting noise properties are shown
with asterisks.  This level of masking removes 16\% of the pixels from
the roughly 38.5 million pixels.  Increasing the size of the masks to
1.5 (2.0) times the SExtractor determined object radii produces noise
properties shown with a square (triangle) symbol.  These increases in
mask size increase the fraction of masked pixels to 62\% and 75\%
respectively.  Further increases in object mask radii only make
inconsequential changes to the noise properties, indicating that at
twice the object radii is a good way to remove all of the flux from
the galaxies.  We also perform the test of using a mask that is the
composite mask of 3.6 \& 4.5 \micron~ object shapes.  This slightly
changes the levels of the masks in this medium scale region in the
same way that changing the individual mask size affects noise
measurement.  Because we do see a change in noise level with changing
mask size, we assume the discrepancy in this binning length regime is
dominated by noise from the wings of galaxies which are improperly
masked.

Even after increasing the mask sizes there remains extra sources of
noise preventing us from achieving ultra-low surface brightness
limits.  There appears to be a floor to the noise at roughly 0.0005
MJy/sr/pixel at 3.6 \micron~ and 0.0008 MJy/sr/pixel at 4.5 \micron.

\subsection{Large Scales}

We expect some of the large scale noise to be caused by the mapping
pattern used for the observations.  Our coverage map is not completely
uniform, so on scales of roughly half a field of view there are
differences in the total exposure time and hence the total number of
electrons detected.  While this plot is shown in MJy/sr, we have
performed the test of converting it to electrons using the coverage
map.  With this test we find an inconsequential change in the noise
properties on large scales, certainly nothing that brings it in line
with the expected linear relation.

There are remaining sources of noise on these large scales, both
instrumental and astronomical, which are very hard to disentangle.
Uncertainties in the flat-fielding, and in our method of removing the
first frame effect are two instrumental effects which are contributing
to the noise on large scales.  We expect the first frame effect to
have a column-wise dependence that requires special calibration data
to measure.  There is potentially also noise due to the ICL being blue
in the IRAC bands while the zodiacal light from which the flats are
made is red in the IRAC bands.  Astrophysically, there is real
structure in the zodiacal light and Galactic cirrus.  There is also
documented diffuse intracluster light in the Virgo cluster itself and
a small signal from the extragalactic background light which are both
adding signal at low levels.  Differentiating between these sources of
error is no easy task (especially given the lack of a close-able
shutter on IRAC).  It is beyond the scope of this paper to discuss the
exact contributions to the noise from these sources.  Instead, we take
this low surface brightness floor simply as an accurate measure of the
noise on large scales.  This floor level is very similar to the quoted
extragalactic background light level in published IRAC work on that
topic \citep[See][ also for a discussion of the errors in each of the
potential instrumental and astronomical sources ]{kashlinsky2005}.

\section{Results \& Discussion}
\label{results}

Before discovering the noise floor, our program was designed to reach
depths comparable to the LSB optical detections in this field with
many resolution elements per plume.  While we are limited by this
floor, there is still interesting science to be done with the
IRAC data.  A first-look through the images reveals $\sim100,000$
sources in both channels.

\subsection{A New Galaxy Cluster Behind Virgo}
\label{new_cluster}
\begin{figure}
\begin{center}
\includegraphics[angle=90, scale=0.4]{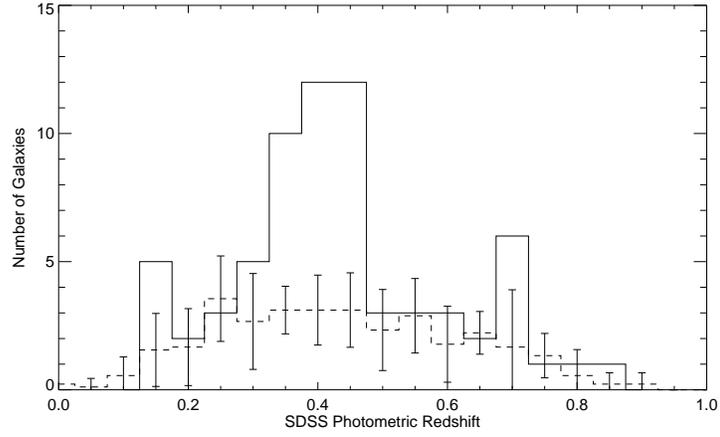}
\end{center}
\caption[backcluster]{Redshift distribution of a galaxy cluster behind
  the Virgo cluster.  The solid line shows the redshift distribution of
  all SDSS detected galaxies with photometric redshifts (and errors on
  those redshifts below 0.2) within 0.04 degrees of the center of the
  new cluster.  The dashed line shows the mean redshift distribution
  in the same size regions near to this new cluster.  The error bars
  are one standard deviation on the mean of those background
  distributions. }
\label{fig:backcluster}
\epsscale{1}
\end{figure}

Among the $\sim 100,000$ detected sources, we have discovered a
potential galaxy cluster behind the Virgo cluster at a redshift of z
$\simeq 0.4$.  This candidate cluster jumps out as a grouping of
galaxies in the IRAC data which is not noticeable in the optical
imaging.  We use SDSS catalog photometric redshifts to investigate
this apparent over-density of sources in the IRAC images at 186.76731,
+13.1463 (J2000).  Figure \ref{fig:backcluster} shows the redshift
distribution in a circle of radius 0.04 degrees centered on the
candidate cluster.  This distribution is compared to the mean redshift
distribution in nine similarly defined regions within the same area on
the sky.  The standard deviation on the comparison distributions is
also shown.  The apparent over-densities at z $\simeq$
0.15 and 0.7 are less than two sigma. The peak at z $\simeq 0.4$ is highly
significant at 11 sigma.  This redshift peak is not detected in any of the
published SDSS cluster catalogs \citep{koester2007,
  miller2005,wen2009}.  The only coincident objects listed in NED are
the galaxies themselves and one ROSAT detected X-ray point source.
This source requires further follow-up for confirmation, including
photo-z determination with the IRAC datapoints included, and
spectroscopic confirmation.

\subsection{Intracluster Plumes}
\label{plumes}
\begin{figure}
\begin{center}
\plotone{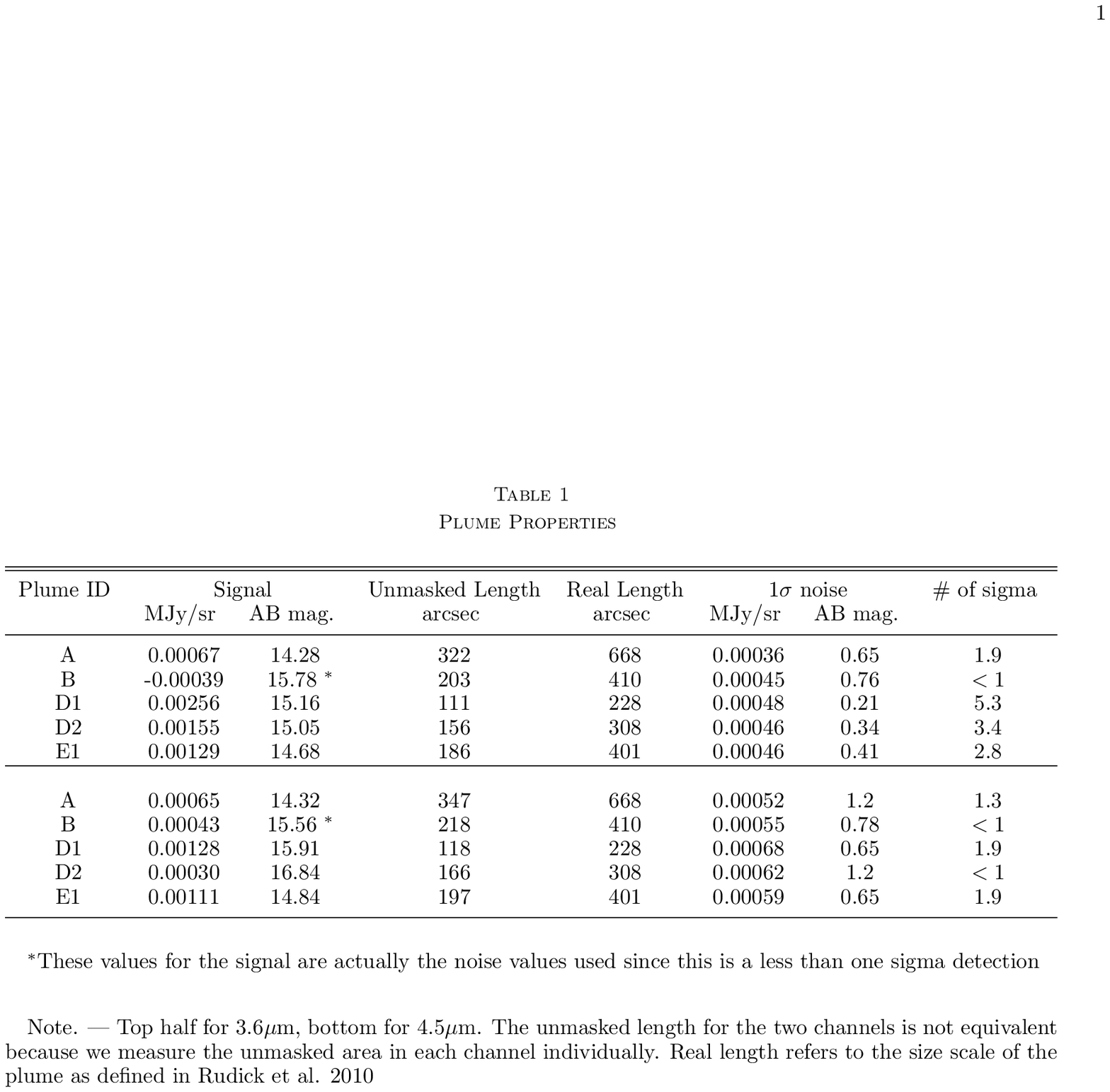}
\end{center}
\label{tab:plumes}

\end{figure}

Our goal in this paper is to detect and characterize the optically
detected plumes in the region between M86 and M87. Our {\it Spitzer}
images overlap with four optically detected regions identified by
\citet{rudick2010} and shown in Figure \ref{fig:ch1}.  These regions
include three plumes(A, B, \& D) and one region in the outer halo of
M87 (E1). Fluxes are measured by summing pixels inside of the four
regions on our final reduced mosaic.  As with the noise measurement,
all objects on the final mosaic are masked to a radius of twice their
detected radii (see \S\ref{noise_medium} ).  Surface brightness and
noise measurements for these regions are shown in Table
\ref{tab:plumes}.  At 3.6 \micron, plume D is a five sigma, secure
detection, while plumes A \& B are not solid detections (1.9 and less
than 1 sigma respectively).  Region E1 is detected at nearly three
sigma.  At 4.5 \micron, no regions are detected with high
significance.

Plume D is an interesting source, likely Galactic in origin
\citep{cortese2010}, which will be discussed in a future paper.  We
focus here on plumes A \& B and region E1.  Plumes A \& B are not
detected in the 100 \micron~ IRAS images.  A non-detection implies
that these plumes are extragalactic, although it is still possible
that they are Galactic features below the sensitivity of IRAS.

\subsubsection{Stellar Mass \& Age}

We estimate both the stellar mass and age of the intracluster plumes
by using our the individual facets of our optical and infrared data
where they have the most power.  There are three important pieces of
information we can use. (1) B-V color puts a constraint on the age of
the population.  This is because the slope of the Wien side of the
blackbody is more sensitive to age variations than the Rayleigh-Jeans.
(2) IRAC flux upper limits put constraints on the upper limit of the
stellar mass.  The power of the infrared is in using the 3.6 \micron~
flux to study mass. This is because most of the mass in an old
population is emitting near the peak of the stellar distribution,
whereas the young massive stars that emit heavily in the optical have
faded away.  This leads to large variations in the optical M/L ratio
as a function of age, especially for old populations, while the
infrared M/L is relatively constant over the same age range.
Additionally, the infrared is a good place to study mass because it is
not affected by dust extinction, as the optical is, and we do not
expect hot dust emission to be affecting the IRAC fluxes due to the
predicted old age of the population. Lastly, (3) Optical fluxes put
constraints on the lower limit to the mass.  While the optical is not
the best place to determine the stellar mass because of the reasons
listed above, given the age of the stellar population from the colors,
and assuming a dustless model, we can put a lower limit on the stellar
mass.  Any stellar population model that falls below our optical data
points is ruled out by the optical detections.

\begin{figure}
\begin{center}
\includegraphics[scale=0.5]{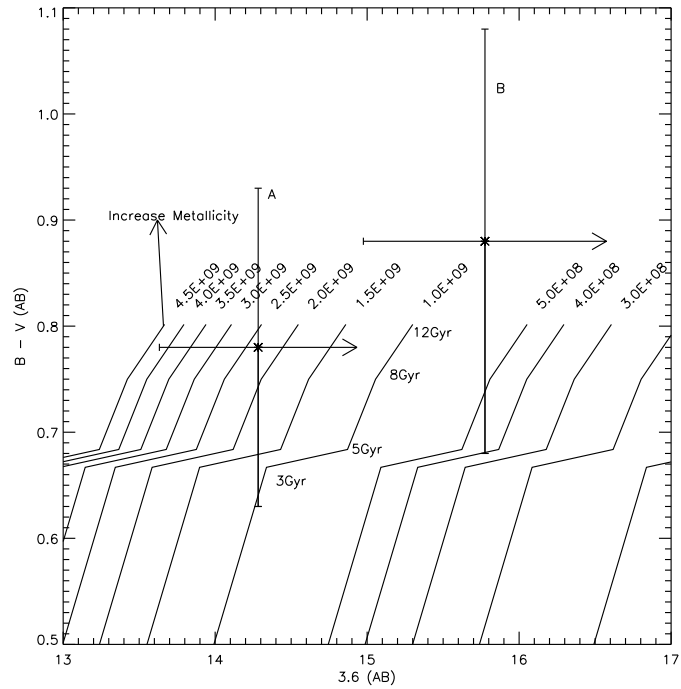}
\end{center}
\caption[mass]{Color magnitude diagram of plumes A \& B for stellar
  mass and age determination.  Points are the data for both plumes
  including one sigma error bars or upper limits.  Solid lines show constant mass tracks as a
  function of age.  Masses are listed at the oldest point on each
  line.  Ages are shown for the 1.0E9 mass track.  An arrow indicates
  the direction that the mass tracks would move if the metallicity
  were increased to the solar value.}
\label{fig:mass}
\epsscale{1}
\end{figure}

With these three pieces of information we constrain both age and
stellar mass of the populations for plumes A \& B starting with a
color magnitude diagram of B - V vs our upper limits on 3.6 \micron~
flux (Figure \ref{fig:mass} ).  The 3.6 \micron~ fluxes for plumes A
\& B are both considered upper limits to the actual fluxes because
they are not more than three sigma above the noise.  For the case of
plume B, the flux measured is less than one sigma, so we use the noise
value itself as the observed flux density. For comparison, we use a family of
\citet{bruzual2003} models of a sub-solar metallicty ([Fe/H] = -0.33)
simple stellar populations as elliptical galaxy templates.
\citet{williams2007} work on {\it HST} ACS data of individual ICL
stars in Virgo suggest they have sub-solar metallicity.  The only
variables in the models are age (1, 3, 5, 8, and 12 Gyr; our error
bars do not warrant higher resolution) and mass($1\times 10^8$ to
$4.5\times 10^9$ \msun).  We choose to use this elliptical template
since ICL sources have been shown to be optically red, both from
observations \citep{krick2006, rudick2010} and from theoretical work
\citep{willman2004,murante2004}.

We first examine age from the B-V color.  We expand the work of
\citet{rudick2010} by showing the full range of model ages consistent
with their measured B - V colors for the plumes A \& B.  Plume A is
consistent with a population older than 3Gyr, while plume B is
consistent with a population older than 5Gyr.  Age determinations from
the optical colors measured are also affected by the metallicity of
the plumes.  In Figure \ref{fig:mass} we show the direction that the
mass curves would move if the metallicity were increased by a factor
of 2.5 to the solar value.  If the metallicity is larger than our
assumed value, then the plumes are younger than we predict, and
viceversa.

We next examine an upper limit to the stellar mass from the 3.6
\micron~ fluxes.  We quote here a range of masses which fit inside the
error bars of the ICL plumes, they represent a conservative upper
limit to the mass of the plumes.  The upper limit to the mass of plume
A is $1.1\times 10^9 - 4.5\times 10^9$ \msun. The upper limit to the
mass of plume B is $3.0\times 10^8 - 1.5\times 10^9$ \msun.  We have
also checked that the \citet{maraston2005} models give consistant
estimates.  

Lower limits to the mass are inferred from to combination of the age
model with the optical fluxes.  We again look at the simple stellar
population model described above.  We choose the lowest age that fits
the B-V color (3Gyr for Plume A, and 5Gyr for plume B), and determine
which masses are ruled out by the detections in the optical bands
which gives a lower limit to the masses of the plumes.  Using a model
without dust here will give the lowest lower limit because assuming
dust will only increase the inherent optical fluxes and thus increase
the mass estimates.  The lower limit to the mass of plume A is
$5.5\times 10^8$ and B is $2.1\times 10^8$.  In conclusion, assuming
the highest upper limit and the lowest lower limit we find the masses
of plume A and B to be between $5.5\times 10^8 - 4.5\times 10^9$ and
$2.1\times 10^8 - 1.5\times 10^9$ \msun respectively.  These are the
first mass measurements of intracluster light plumes.

The masses we find for plumes A \& B are consistent with those found
in the simulations of \citet{rudick2009}.  Other theoretical studies of ICL
production do not discuss plume/stream properties
\citep{henriques2010,murante2007,sommer-larsen2005,willman2004}

For comparison we calculate the total stellar mass in Virgo from the
spectroscopic luminosity function of \citet{rines2008}.  We integrate
under the Schechter function fit to their luminosity function from
$M_r = -11$ to $M_r = -23$.  This gives us a total luminosity of
$2.9\times10^{11}$ \lsun.  If we assume a mass to light ratio for
this cluster of 500 \citep{schindler1999}, we find a total cluster
mass of $1.5\times 10^{14}$ \msun.  This is in the same range as found
by \citet{schindler1999} based on X-ray and optical observations who
find a total mass of both the M87 and M49 groups of $3\times 10^{14}
\pm0.6$ \msun.

In total, if plumes A \& B are extragalactic, and if they were the
only intracluster light in the Virgo cluster they would represent at
most a couple percent of the total stellar mass in the cluster.  Our
{\it Spitzer} mosaic covers roughly half of the area around M87 and
half of the area around M86/M84, so if the density of plumes/streams
is constant in the region around the center of the cluster we could
imagine that at most a few percent of the cluster mass is in ICL
streams at this particular point in the Virgo cluster's
evolution. Therefore, at any given time in the cluster's history, the
plumes do not account for a large fraction of the cluster mass; but we
expect plume generation and destruction to be ongoing in the cluster.

Simulations suggest that these plumes last for less than one Gyr, or
roughly 1-2 crossing times \citep{rudick2009}.  Assuming the cluster
has been forming for roughly ten Gyr, and assuming a constant rate of
plume formation, we naively expect to find a few tens of percent of
the total stellar mass be formed in plumes over the lifetime of this
cluster.  This is consistent with total ICL fractions as measured from
the diffuse ICL component of 10-40\% of the total cluster light
\citep{krick2007,gonzalez2005,feldmeier2004}.  This suggests that
bright plumes can account, order of magnitude, for all of the diffuse
ICL as measured.  We then don't need to rely on ram pressure stripping
(RPS), or other gas dynamical effects for input to the diffuse ICL.  Although these
processes do occur, they might not be the dominant input mechanism to
the ICL.  This could explain why intracluster light appears red,
without any evidence for {\it in situ} star formation which you might
expect to see if RPS were the dominant input mechanism. While we can't
rule out the other enrichment mechanisms, our data is consistent with
gravitational processes which create bright plumes as the main
enrichment mechanism to the ICL.

\subsection{M87 Halo}
\label{SED}

We combine our infrared data on the halo of M87 to unprecedented radii
with both optical and infrared archival data of the center of
M87.  The infrared data will allow us to understand if there is a dust
component in M87 which may be effecting the optical estimates of age.
From this age gradient information we hope to learn about the
potential formation mechanisms of this giant elliptical.

\begin{figure}
\begin{center}
\includegraphics[angle=90, scale=0.5]{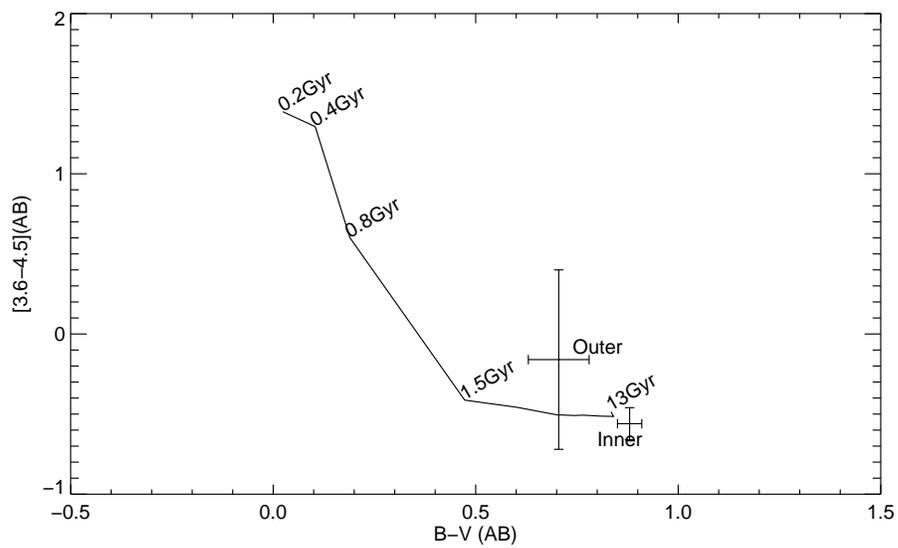}
\end{center}
\caption[grasil]{Color-color plot of the M87 halo.  The points
  indicate the colors of the inner region and outer halo of M87.  The
  line is a Grasil model of elliptical galaxy evolution as a function
  of age.  The outer halo is consistent with having the same color as
  the inner region of the galaxy.}
\label{fig:grasil}
\epsscale{1}
\end{figure}

To learn about the stellar populations in the outer halo of M87, we
examine the optical and infrared color profile to large radii in
Figure \ref{fig:grasil}.  The outer halo of M87 is defined by region
E1, shown in Figure \ref{fig:ch1}, 160 kpc (2008 arcseconds) from the
center of M87.  This is four times the radius analyzed by a recent
multi-band optical study by \citet{liu2005} (even though the profiles
in their plots do extend to comparable radii).  Optical measurements
of the same outer halo region comes from \citet{mihos2005} and
\citet{rudick2010}.  For comparison, both infrared and optical
photometry of the central region comes from the literature
\citep{shi2007,zeilinger1993}.  Because M87 has a central jet, we
specifically choose photometry from the literature that carefully
removes the jet component.

For reference, the color-color plot includes a Grasil model
evolutionary track for an elliptical galaxy \citep{silva1998} for ages
ranging from formation to 13 Gyr. These models are different from
standard stellar population models because they include dust in three
environments; interstellar HI clouds heated by the general
interstellar radiation field, star forming molecular clouds and HII
regions, and dusty envelopes of dying stars.  These models are
intended to be elliptical templates.  They have an initial infall of
gas with an exponential decay timescale of 0.1Gyr followed by passive
evolution.  We choose to use models with dust in them because dust is
important in predicting the infrared colors of a younger population.
Note that these models show infrared colors becoming bluer with age,
which is opposite the color trend of the optical data.  When we
discuss trends below we will use ``blue'' in the sense of the optical
data.

We use our data to study the existence of dust in clusters.  Dust
grains should be present in the ICM because we know that metals, which
are often depleted onto grains, escape galaxies by ram pressure
stripping, supernova explosions, and tidal interactions It is not
clear how long dust is able to survive in the ICM due to the
destructive effects of sputtering off of the hot ICM
\citep{yamada2005}. The direct detection of dust in the far-infrared
has so far not been possible due to the large background fluctuations
and limited telescope resolution, making it difficult to determine if
signal is from the ICM or galaxies \citep{bai2007, kitayama2009,
  montier2005, giard2008}. However, there is substantial evidence for
intracluster dust. Many low-redshift cluster galaxies exhibit
significant amounts of extraplanar 8 μm emission (e.g. NGC 4501 \& NGC
4522) which is thought to be the result of polycyclic aromatic
hydrocarbons (PAHs) begin pulled out of the galaxy by ram pressure
stripping. Further evidence comes from \citet{girardi1992}, who find
that asymmetric redshift distributions in nearby groups implies
extinction consistent with dust obscuration. Additionally,
\citet{maoz1995, McGee2010} find that high-redshift objects behind
  clusters show large visual extinctions.

To get a broad idea of the presence of dust, we compare age
predictions from the optical and infrared data.  Age predictions from
our infrared colors are consistent with age predictions from the
optical data within $2\sigma$ which implies broadly that dust is not
effecting the optical colors.  Dust can cause extinction at optical
wavelengths.  With a low surface brightness measurement, there is no
good way of ruling out dust extinction without infrared data which can
easily penetrate the dust.  3.6 \& 4.5 \micron~ observations provide
this without being effected by hot dust emission.

The second thing to notice is that any interpretation of a color
change in the profile from the inner region to the outer halo at 160
kpc is a two sigma measurement.  Both a flat color profile, and a
slightly blueing optical profile are consistent with our measurements
within two sigma.  In other words, our data do not place strong
constraints on the color profile of the halo of M87.

Optical work on the color profile at large radii has been interpreted
as having a blue trend attributed to younger ages with radius and/or a
potential metallicity
gradient\citet{zeilinger1993,liu2005,rudick2010}.  We note that the
detection of a blue trend in the optical work at large radii is also a
low sigma measurement.

A flat or blue optical gradient in the outer halo is also consistent
with simulations.  Specifically, ICL simulations form the ICL stars at
redshifts greater than 1.5 \citep{murante2004, sommer-larsen2005},
which would therefore make them red at low redshift, causing the outer
halos of galaxies like M87 to also be red like their centers.
Numerical simulations of two clusters in \citet{sommer-larsen2005}
predict a flat to slightly blue-ward profile in optical colors from a
radius of 10kpc to 1Mpc from the center of the brightest cluster
galaxy (BCG).  They find a decrease in metallicity with radius causing
the blue-ward trend.  According to this work then, differentiation
between a blue or flat optical gradient in the outer halo would give
us information on the metallicity of the ICL population.

Since this color profile extends to the largest radii yet discussed,
we are likely sampling the ICL population. The relatively old stellar
ages in both the inner and outer profile to the limits of our observations
suggests that the halo of M87 has not been infused with a young
population in the last few Gyr.  A young ICL population could come
from either {\it in situ} formation triggered by interactions with other
galaxies or the cluster potential.  Or it could come from the
stripping of already young stars away from an infalling galaxy.  The red colors of the
 outer halo rule out these cases for this massive galaxy in the deep potential
well of center of the Virgo cluster.
 
We hesitate to say that the flat color gradient in the outer halo is a
byproduct of monolithic collapse.  Because the surface brightness
profile shows a distinct, outer ICL component \citep{liu2005}, it is
likely that the outer component has a different formation mechanism
than the inner deVaucouleurs profile.  However, our findings are
consistent with the outer profile being composed of stars similarly
old as those in the center of the galaxy.


\section{Conclusion}
\label{conclusion}

In summary, we describe {\it Spitzer} IRAC imaging of the central 0.77
square degrees of the Virgo cluster with specific attention paid to
the low surface brightness features in this region.  This is the first
intracluster light study in the infrared.  We reach a limiting SB of
$3.6\times 10^(-4)$ MJy/sr for the largest regions. We don't detect
the intracluster light plumes as seen in the optical. From upper
limits on their brightnesses we present an optical and infrared color
magnitude diagram and the first ever measurement of the mass of an
intracluster light plume.  We overlay that plot with a family of model
elliptical galaxies which allows us to put an upper limit on the mass
and age of the plumes.  Then optical colors are used to constrain the
lower limit to the masses.  We do this for two plumes finding masses
in the range of $5.5\times 10^8 - 4.5\times 10^9$ and $2.1\times 10^8
- 1.5\times 10^9$ \msun for the two plumes for which we have coverage.
These are consistent with streams formed in simulations.  In total,
these types of plumes, assumed constant formation rate and summing
over the lifetime of the cluster, are massive enough to account for
the entire diffuse ICL population.

Secondly we look at the color profile of M87.  Age predictions from
our infrared colors are consistent with age predictions from the
optical data which implies that dust is not effecting the optical
colors.  Comparing our infrared data on the outer halo of M87 with
literature values for the inner region and optical profile, we find
that the M87 color profile at large radius is either flat or becomes
slightly more optically blue with radius.  A blue profile would indicate younger
or lower metallicity stars at larger radii.  Although a flat color
profile is strictly consistent with monolithic collapse models, M87 is
known to have a diffuse outer ICL component with a different radial
profile, implying a different formation mechanism compared to the rest
of M87.

Lastly, we comment on the intricacies of the IRAC instrument, and the
best methods to observe LSB features.  LSB work with
IRAC requires careful attention to details of the instrument and of
the astrophysical background.  Due to instrumental effects, large
scale binning of the data does not allow the observer to reach deeper
surface brightness detection limits as expected.  Increased exposure time does afford
the deepest possible surface brightnesses, as described in the
appendix.


\acknowledgments

We thank Sabrina Stierwalt and the IRAC instrument support team for
useful discussions.  This research has made use of data from the
Infrared Processing and Analysis Center/California Institute of
Technology, funded by the National Aeronautics and Space
Administration and the National Science Foundation.  This work was
based on observations obtained with the {\it Spitzer} Space Telescope,
which is operated by the Jet Propulsion Laboratory, California
Institute of Technology under a contract with NASA. This research has
made use of the NASA/IPAC Extragalactic Database (NED) which is
operated by the Jet Propulsion Laboratory, California Institute of
Technology, under contract with the National Aeronautics and Space
Administration.

{\it Facilities:} \facility{Spitzer (IRAC)}

\bibliography{cluster}  

\begin{thebibliography}{62}
\expandafter\ifx\csname natexlab\endcsname\relax\def\natexlab#1{#1}\fi

\bibitem[{{Abadi} {et~al.}(1999){Abadi}, {Moore}, \& {Bower}}]{abadi1999}
{Abadi}, M.~G., {Moore}, B., \& {Bower}, R.~G. 1999, \mnras, 308, 947

\bibitem[{{Abramson} {et~al.}(2011){Abramson}, {Kenney}, {Crowl}, {Chung}, {van
  Gorkom}, {Vollmer}, \& {Schiminovich}}]{2011arXiv1101.4066A}
{Abramson}, A., {Kenney}, J.~D.~P., {Crowl}, H.~H., {Chung}, A., {van Gorkom},
  J.~H., {Vollmer}, B., \& {Schiminovich}, D. 2011, ArXiv e-prints

\bibitem[{{Bai} {et~al.}(2007){Bai}, {Rieke}, \& {Rieke}}]{bai2007}
{Bai}, L., {Rieke}, G.~H., \& {Rieke}, M.~J. 2007, \apjl, 668, L5

\bibitem[{{Bekki}(1999)}]{bekki1999}
{Bekki}, K. 1999, \apjl, 510, L15

\bibitem[{{Bertin} \& {Arnouts}(1996)}]{bertin1996}
{Bertin}, E., \& {Arnouts}, S. 1996, \aaps, 117, 393

\bibitem[{{Bruzual} \& {Charlot}(2003)}]{bruzual2003}
{Bruzual}, G., \& {Charlot}, S. 2003, \mnras, 344, 1000

\bibitem[{{Butcher} \& {Oemler}(1978)}]{butcher1978}
{Butcher}, H., \& {Oemler}, Jr., A. 1978, \apj, 219, 18

\bibitem[{{Butcher} \& {Oemler}(1984)}]{butcher1984}
---. 1984, \apj, 285, 426

\bibitem[{{Byrd} \& {Valtonen}(1990)}]{Byrd1990}
{Byrd}, G., \& {Valtonen}, M. 1990, \apj, 350, 89

\bibitem[{{Cortese} {et~al.}(2010){Cortese}, {Bendo}, {Isaak}, {Davies}, \&
  {Kent}}]{cortese2010}
{Cortese}, L., {Bendo}, G.~J., {Isaak}, K.~G., {Davies}, J.~I., \& {Kent},
  B.~R. 2010, \mnras, 403, L26

\bibitem[{{Desai} {et~al.}(2007){Desai}, {Dalcanton}, {Arag{\'o}n-Salamanca},
  {Jablonka}, {Poggianti}, {Gogarten}, {Simard}, {Milvang-Jensen}, {Rudnick},
  {Zaritsky}, {Clowe}, {Halliday}, {Pell{\'o}}, {Saglia}, \&
  {White}}]{desai2007}
{Desai}, V. {et~al.} 2007, \apj, 660, 1151

\bibitem[{{Dressler} {et~al.}(1997){Dressler}, {Oemler}, {Couch}, {Smail},
  {Ellis}, {Barger}, {Butcher}, {Poggianti}, \& {Sharples}}]{dressler1997}
{Dressler}, A. {et~al.} 1997, \apj, 490, 577

\bibitem[{{Fazio} {et~al.}(2004){Fazio}, {Hora}, {Allen}, {Ashby}, {Barmby},
  {Deutsch}, {Huang}, {Kleiner}, {Marengo}, {Megeath}, {Melnick}, {Pahre},
  {Patten}, {Polizotti}, {Smith}, {Taylor}, {Wang}, {Willner}, {Hoffmann},
  {Pipher}, {Forrest}, {McMurty}, {McCreight}, {McKelvey}, {McMurray}, {Koch},
  {Moseley}, {Arendt}, {Mentzell}, {Marx}, {Losch}, {Mayman}, {Eichhorn},
  {Krebs}, {Jhabvala}, {Gezari}, {Fixsen}, {Flores}, {Shakoorzadeh}, {Jungo},
  {Hakun}, {Workman}, {Karpati}, {Kichak}, {Whitley}, {Mann}, {Tollestrup},
  {Eisenhardt}, {Stern}, {Gorjian}, {Bhattacharya}, {Carey}, {Nelson},
  {Glaccum}, {Lacy}, {Lowrance}, {Laine}, {Reach}, {Stauffer}, {Surace},
  {Wilson}, {Wright}, {Hoffman}, {Domingo}, \& {Cohen}}]{fazio2004}
{Fazio}, G.~G. {et~al.} 2004, \apjs, 154, 10

\bibitem[{{Feldmeier} {et~al.}(2004){Feldmeier}, {Mihos}, {Morrison},
  {Harding}, {Kaib}, \& {Dubinski}}]{feldmeier2004}
{Feldmeier}, J.~J., {Mihos}, J.~C., {Morrison}, H.~L., {Harding}, P., {Kaib},
  N., \& {Dubinski}, J. 2004, \apj, 609, 617

\bibitem[{{Giard} {et~al.}(2008){Giard}, {Montier}, {Pointecouteau}, \&
  {Simmat}}]{giard2008}
{Giard}, M., {Montier}, L., {Pointecouteau}, E., \& {Simmat}, E. 2008, \aap,
  490, 547

\bibitem[{{Girardi} {et~al.}(1992){Girardi}, {Mezzetti}, {Giuricin}, \&
  {Mardirossian}}]{girardi1992}
{Girardi}, M., {Mezzetti}, M., {Giuricin}, G., \& {Mardirossian}, F. 1992,
  \apj, 394, 442

\bibitem[{{Gonzalez} {et~al.}(2005){Gonzalez}, {Zabludoff}, \&
  {Zaritsky}}]{gonzalez2005}
{Gonzalez}, A.~H., {Zabludoff}, A.~I., \& {Zaritsky}, D. 2005, \apj, 618, 195

\bibitem[{{Gunn} \& {Gott}(1972)}]{gunn1972}
{Gunn}, J.~E., \& {Gott}, J.~R.~I. 1972, \apj, 176, 1

\bibitem[{{Henriksen} \& {Byrd}(1996)}]{henriksen1996}
{Henriksen}, M., \& {Byrd}, G. 1996, \apj, 459, 82

\bibitem[{{Henriques} \& {Thomas}(2010)}]{henriques2010}
{Henriques}, B.~M.~B., \& {Thomas}, P.~A. 2010, \mnras, 403, 768

\bibitem[{{Janowiecki} {et~al.}(2010){Janowiecki}, {Mihos}, {Harding},
  {Feldmeier}, {Rudick}, \& {Morrison}}]{janowiecki2010}
{Janowiecki}, S., {Mihos}, J.~C., {Harding}, P., {Feldmeier}, J.~J., {Rudick},
  C., \& {Morrison}, H. 2010, \apj, 715, 972

\bibitem[{{Kashlinsky} {et~al.}(2005){Kashlinsky}, {Arendt}, {Mather}, \&
  {Moseley}}]{kashlinsky2005}
{Kashlinsky}, A., {Arendt}, R.~G., {Mather}, J., \& {Moseley}, S.~H. 2005,
  \nat, 438, 45

\bibitem[{{Kelsall} {et~al.}(1998){Kelsall}, {Weiland}, {Franz}, {Reach},
  {Arendt}, {Dwek}, {Freudenreich}, {Hauser}, {Moseley}, {Odegard},
  {Silverberg}, \& {Wright}}]{kelsall1998}
{Kelsall}, T. {et~al.} 1998, \apj, 508, 44

\bibitem[{{Kenney} {et~al.}(2008){Kenney}, {Tal}, {Crowl}, {Feldmeier}, \&
  {Jacoby}}]{kenney2008}
{Kenney}, J.~D.~P., {Tal}, T., {Crowl}, H.~H., {Feldmeier}, J., \& {Jacoby},
  G.~H. 2008, \apjl, 687, L69

\bibitem[{{Kitayama} {et~al.}(2009){Kitayama}, {Ito}, {Okada}, {Kaneda},
  {Takahashi}, {Ota}, {Onaka}, {Tajiri}, {Nagata}, \& {Yamada}}]{kitayama2009}
{Kitayama}, T. {et~al.} 2009, \apj, 695, 1191

\bibitem[{{Koester} {et~al.}(2007){Koester}, {McKay}, {Annis}, {Wechsler},
  {Evrard}, {Bleem}, {Becker}, {Johnston}, {Sheldon}, {Nichol}, {Miller},
  {Scranton}, {Bahcall}, {Barentine}, {Brewington}, {Brinkmann}, {Harvanek},
  {Kleinman}, {Krzesinski}, {Long}, {Nitta}, {Schneider}, {Sneddin}, {Voges},
  \& {York}}]{koester2007}
{Koester}, B.~P. {et~al.} 2007, \apj, 660, 239

\bibitem[{{Kormendy} {et~al.}(2009){Kormendy}, {Fisher}, {Cornell}, \&
  {Bender}}]{kormendy2009}
{Kormendy}, J., {Fisher}, D.~B., {Cornell}, M.~E., \& {Bender}, R. 2009, \apjs,
  182, 216

\bibitem[{{Krick} \& {Bernstein}(2007)}]{krick2007}
{Krick}, J.~E., \& {Bernstein}, R.~A. 2007, \aj, 134, 466

\bibitem[{{Krick} {et~al.}(2006){Krick}, {Bernstein}, \&
  {Pimbblet}}]{krick2006}
{Krick}, J.~E., {Bernstein}, R.~A., \& {Pimbblet}, K.~A. 2006, \aj, 131, 168

\bibitem[{{Krick} {et~al.}(2009){Krick}, {Surace}, {Thompson}, {Ashby}, {Hora},
  {Gorjian}, \& {Yan}}]{krick2009}
{Krick}, J.~E., {Surace}, J.~A., {Thompson}, D., {Ashby}, M.~L.~N., {Hora},
  J.~L., {Gorjian}, V., \& {Yan}, L. 2009, \apj, 700, 123

\bibitem[{{Larson} {et~al.}(1980){Larson}, {Tinsley}, \&
  {Caldwell}}]{Larson1980}
{Larson}, R.~B., {Tinsley}, B.~M., \& {Caldwell}, C.~N. 1980, \apj, 237, 692

\bibitem[{{Liu} {et~al.}(2005){Liu}, {Zhou}, {Ma}, {Wu}, {Yang}, {Li}, \&
  {Chen}}]{liu2005}
{Liu}, Y., {Zhou}, X., {Ma}, J., {Wu}, H., {Yang}, Y., {Li}, J., \& {Chen}, J.
  2005, \aj, 129, 2628

\bibitem[{{Makovoz} \& {Marleau}(2005)}]{makovoz2005}
{Makovoz}, D., \& {Marleau}, F.~R. 2005, \pasp, 117, 1113

\bibitem[{{Maoz}(1995)}]{maoz1995}
{Maoz}, D. 1995, \apjl, 455, L115+

\bibitem[{{Maraston}(2005)}]{maraston2005}
{Maraston}, C. 2005, \mnras, 362, 799

\bibitem[{{McGee} \& {Balogh}(2010)}]{McGee2010}
{McGee}, S.~L., \& {Balogh}, M.~L. 2010, \mnras, 405, 2069

\bibitem[{{Mihos}(2004)}]{mihos2004}
{Mihos}, J.~C. 2004, in Clusters of Galaxies: Probes of Cosmological Structure
  and Galaxy Evolution, from the Carnegie Observatories Centennial Symposia,
  2004, p. 278., 278--+

\bibitem[{{Mihos} {et~al.}(2005){Mihos}, {Harding}, {Feldmeier}, \&
  {Morrison}}]{mihos2005}
{Mihos}, J.~C., {Harding}, P., {Feldmeier}, J., \& {Morrison}, H. 2005, \apjl,
  631, L41

\bibitem[{{Miller} {et~al.}(2005){Miller}, {Nichol}, {Reichart}, {Wechsler},
  {Evrard}, {Annis}, {McKay}, {Bahcall}, {Bernardi}, {Boehringer}, {Connolly},
  {Goto}, {Kniazev}, {Lamb}, {Postman}, {Schneider}, {Sheth}, \&
  {Voges}}]{miller2005}
{Miller}, C.~J. {et~al.} 2005, \aj, 130, 968

\bibitem[{{Montier} \& {Giard}(2005)}]{montier2005}
{Montier}, L.~A., \& {Giard}, M. 2005, \aap, 439, 35

\bibitem[{{Moore} {et~al.}(1996){Moore}, {Katz}, {Lake}, {Dressler}, \&
  {Oemler}}]{moore1996}
{Moore}, B., {Katz}, N., {Lake}, G., {Dressler}, A., \& {Oemler}, A. 1996,
  \nat, 379, 613

\bibitem[{{Moore} {et~al.}(1998){Moore}, {Lake}, \& {Katz}}]{moore1998}
{Moore}, B., {Lake}, G., \& {Katz}, N. 1998, \apj, 495, 139

\bibitem[{{Murante} {et~al.}(2004){Murante}, {Arnaboldi}, {Gerhard}, {Borgani},
  {Cheng}, {Diaferio}, {Dolag}, {Moscardini}, {Tormen}, {Tornatore}, \&
  {Tozzi}}]{murante2004}
{Murante}, G. {et~al.} 2004, \apjl, 607, L83

\bibitem[{{Murante} {et~al.}(2007){Murante}, {Giovalli}, {Gerhard},
  {Arnaboldi}, {Borgani}, \& {Dolag}}]{murante2007}
{Murante}, G., {Giovalli}, M., {Gerhard}, O., {Arnaboldi}, M., {Borgani}, S.,
  \& {Dolag}, K. 2007, \mnras, 377, 2

\bibitem[{{Murphy} {et~al.}(2009){Murphy}, {Kenney}, {Helou}, {Chung}, \&
  {Howell}}]{murphy2009}
{Murphy}, E.~J., {Kenney}, J.~D.~P., {Helou}, G., {Chung}, A., \& {Howell},
  J.~H. 2009, \apj, 694, 1435

\bibitem[{{Rines} \& {Geller}(2008)}]{rines2008}
{Rines}, K., \& {Geller}, M.~J. 2008, \aj, 135, 1837

\bibitem[{{Rudick} {et~al.}(2009){Rudick}, {Mihos}, {Frey}, \&
  {McBride}}]{rudick2009}
{Rudick}, C.~S., {Mihos}, J.~C., {Frey}, L.~H., \& {McBride}, C.~K. 2009, \apj,
  699, 1518

\bibitem[{{Rudick} {et~al.}(2010){Rudick}, {Mihos}, {Harding}, {Feldmeier},
  {Janowiecki}, \& {Morrison}}]{rudick2010}
{Rudick}, C.~S., {Mihos}, J.~C., {Harding}, P., {Feldmeier}, J.~J.,
  {Janowiecki}, S., \& {Morrison}, H.~L. 2010, \apj, 720, 569

\bibitem[{{Schindler} {et~al.}(1999){Schindler}, {Binggeli}, \&
  {B{\"o}hringer}}]{schindler1999}
{Schindler}, S., {Binggeli}, B., \& {B{\"o}hringer}, H. 1999, \aap, 343, 420

\bibitem[{{Schulz} \& {Struck}(2001)}]{schulz2001}
{Schulz}, S., \& {Struck}, C. 2001, \mnras, 328, 185

\bibitem[{{Shapley} {et~al.}(2001){Shapley}, {Fabbiano}, \&
  {Eskridge}}]{shapley2001}
{Shapley}, A., {Fabbiano}, G., \& {Eskridge}, P.~B. 2001, \apjs, 137, 139

\bibitem[{{Shi} {et~al.}(2007){Shi}, {Rieke}, {Hines}, {Gordon}, \&
  {Egami}}]{shi2007}
{Shi}, Y., {Rieke}, G.~H., {Hines}, D.~C., {Gordon}, K.~D., \& {Egami}, E.
  2007, \apj, 655, 781

\bibitem[{{Silva} {et~al.}(1998){Silva}, {Granato}, {Bressan}, \&
  {Danese}}]{silva1998}
{Silva}, L., {Granato}, G.~L., {Bressan}, A., \& {Danese}, L. 1998, \apj, 509,
  103

\bibitem[{{Sommer-Larsen} {et~al.}(2005){Sommer-Larsen}, {Romeo}, \&
  {Portinari}}]{sommer-larsen2005}
{Sommer-Larsen}, J., {Romeo}, A.~D., \& {Portinari}, L. 2005, \mnras, 39

\bibitem[{{Stevens} {et~al.}(1999){Stevens}, {Acreman}, \&
  {Ponman}}]{stevens1999}
{Stevens}, I.~R., {Acreman}, D.~M., \& {Ponman}, T.~J. 1999, \mnras, 310, 663

\bibitem[{{Vollmer} {et~al.}(2001){Vollmer}, {Cayatte}, {Balkowski}, \&
  {Duschl}}]{vollmer2001}
{Vollmer}, B., {Cayatte}, V., {Balkowski}, C., \& {Duschl}, W.~J. 2001, \apj,
  561, 708

\bibitem[{{Wen} {et~al.}(2009){Wen}, {Han}, \& {Liu}}]{wen2009}
{Wen}, Z.~L., {Han}, J.~L., \& {Liu}, F.~S. 2009, \apjs, 183, 197

\bibitem[{{Williams} {et~al.}(2007){Williams}, {Ciardullo}, {Durrell},
  {Vinciguerra}, {Feldmeier}, {Jacoby}, {Sigurdsson}, {von Hippel}, {Ferguson},
  {Tanvir}, {Arnaboldi}, {Gerhard}, {Aguerri}, \& {Freeman}}]{williams2007}
{Williams}, B.~F. {et~al.} 2007, \apj, 656, 756

\bibitem[{{Willman} {et~al.}(2004){Willman}, {Governato}, {Wadsley}, \&
  {Quinn}}]{willman2004}
{Willman}, B., {Governato}, F., {Wadsley}, J., \& {Quinn}, T. 2004, \mnras,
  355, 159

\bibitem[{{Yamada} \& {Kitayama}(2005)}]{yamada2005}
{Yamada}, K., \& {Kitayama}, T. 2005, \pasj, 57, 611

\bibitem[{{Zeilinger} {et~al.}(1993){Zeilinger}, {Moller}, \&
  {Stiavelli}}]{zeilinger1993}
{Zeilinger}, W.~W., {Moller}, P., \& {Stiavelli}, M. 1993, \mnras, 261, 175

\bibitem[{{Zibetti} {et~al.}(2005){Zibetti}, {White}, {Schneider}, \&
  {Brinkmann}}]{zibetti2005}
{Zibetti}, S., {White}, S.~D.~M., {Schneider}, D.~P., \& {Brinkmann}, J. 2005,
  \mnras, 358, 949

\end{thebibliography}




\clearpage
\appendix
\section{How to Detect Low Surface Brightness Features with IRAC?}
We have shown above that there is a ``floor'' in reaching low surface
brightnesses when binning together many pixels because the noise does
not drop as expected with binning length.  We now investigate if lower
surface brightnesses can be achieved by increasing exposure time.,
i.e. does the noise decrease with the square root of exposure time as
expected?  To look at this we have used IRAC dark field data
\citep{krick2009} as a unique dataset that goes very deep in the same
region of the sky mainly devoid of background.  We don't want any
signal to be in the background (low zodiacal light, low Galactic
diffuse emission, no intracluster light) so we can remove as many
astrophysical sources of noise as possible.

Using all the warm mission dark calibration data to date, we have made
a mosaic of 300 frames, each with 100s exposure time.  Mopex was used
to make this mosaic from the column pulldown corrected CBCDs. We mask
all sources with a SExtractor segmentation image.  We look at all the
unmasked pixels with the same coverage levels, ranging from 12 - 300
images. The noise on the distribution of background values within each
coverage level is the standard deviation of the Gaussian fit to that
distribution. Each distribution has $>750$ pixels in it. For
comparison we have done this same analysis on the dark field mosaics
from the first year of the cryogenic mission.

These plots show that IRAC noise does decrease roughly as expected
with exposure time.  The slight deviation at larger exposure times is
likely caused by the first frame effect and by source wings.  The dark
field data has not been given the special treatment of deriving it's
own first frame effect as discussed for the Virgo dataset above.
Also, the sources have been only simply masked with the SExtractor
segmentation map, and so wings of sources will still add noise to the
images which affects the measurement at lower surface brightness.


\begin{figure}
\epsscale{0.4}
\plotone{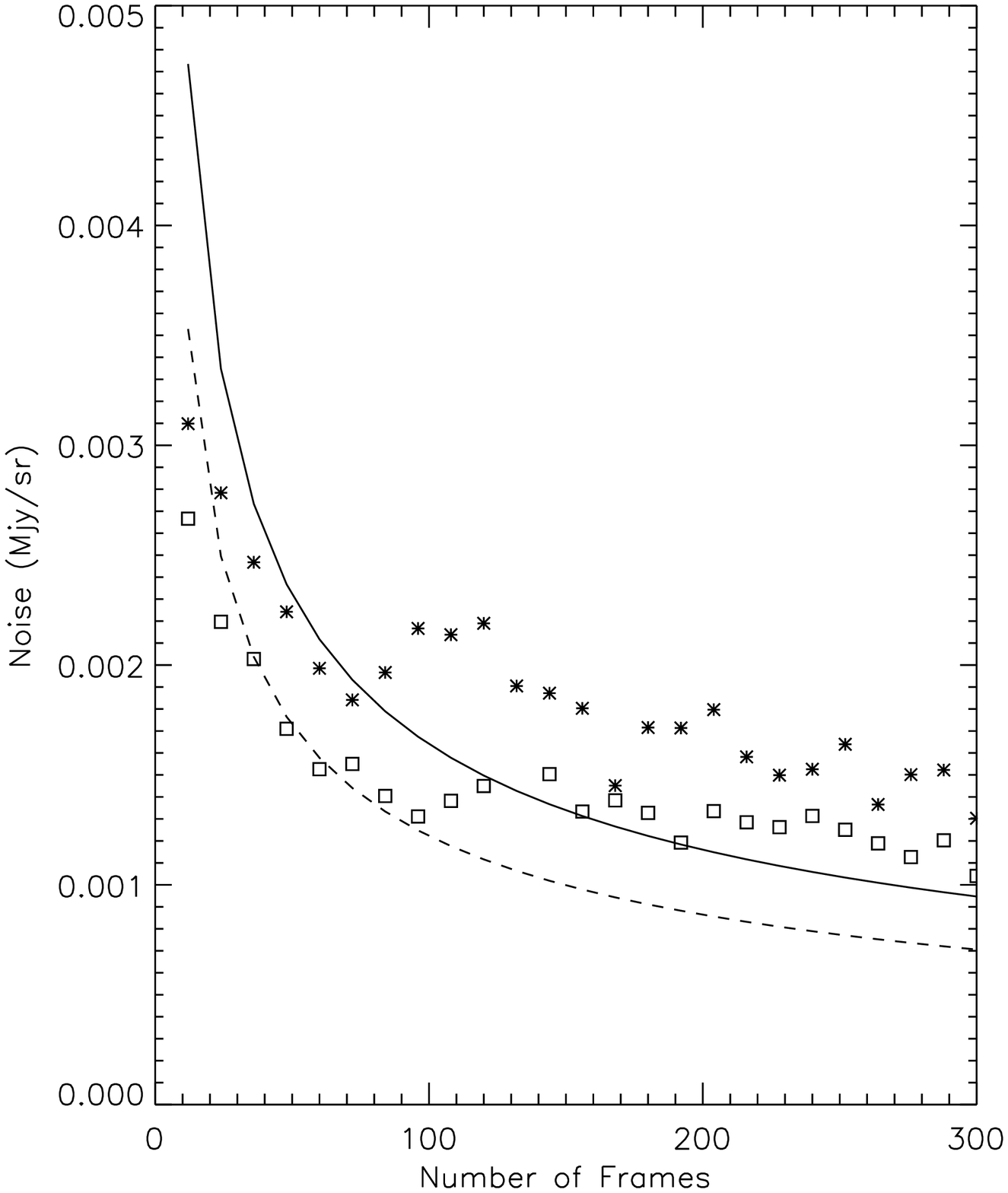}
\plotone{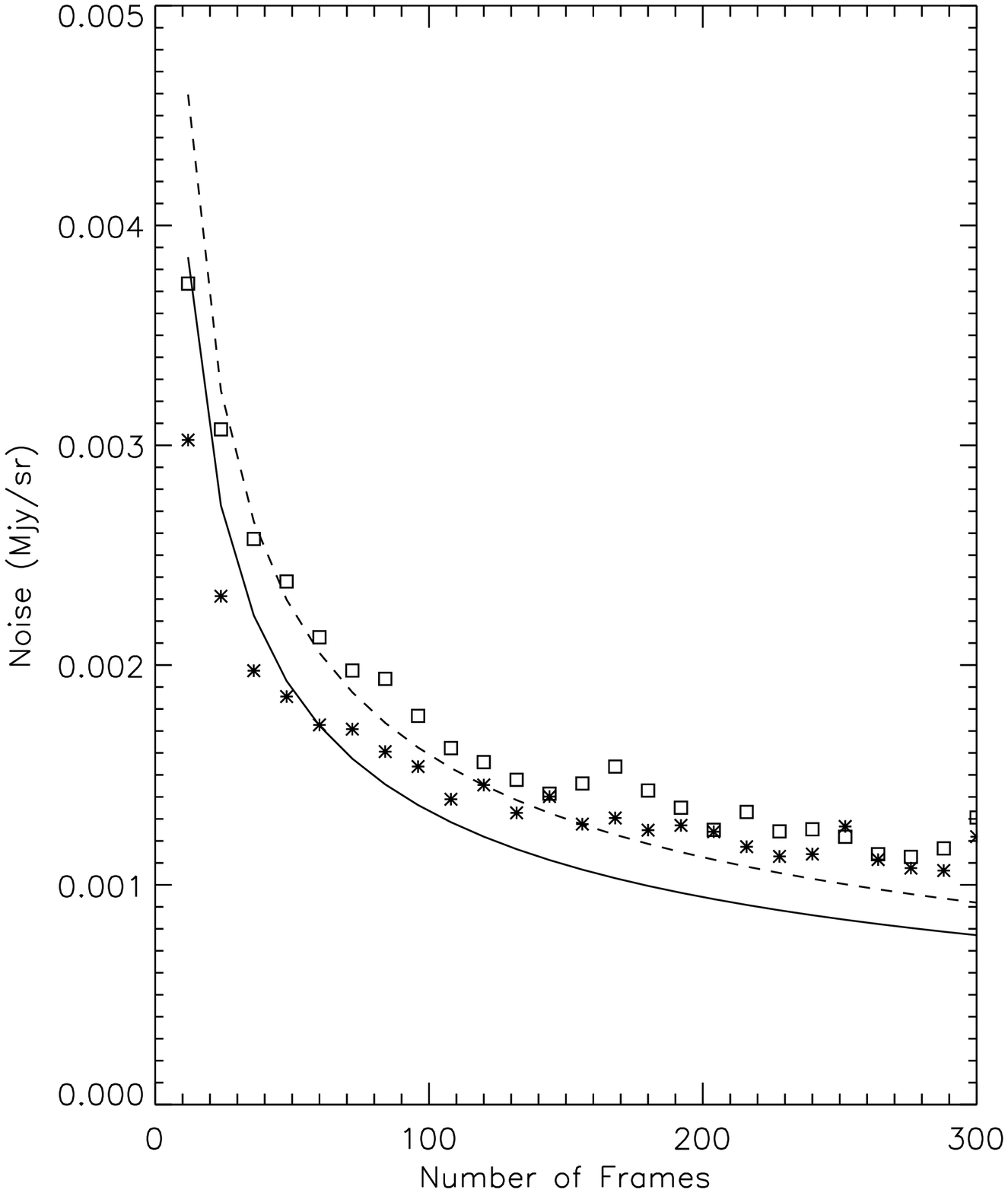}
\caption[psf]{Noise as a function of increasing exposure time; 3.6
  \micron~ on left, 4.5 \micron~ on the right.  Stars and solid lines are the
  warm data and fitted square root function.  Squares and dashed line
  are the same for the cryogenic mission.}
\label{fig:noise_exptime}
\epsscale{1}
\end{figure}


\end{document}